\documentclass{aastex6}

\usepackage{natbib}
\usepackage{graphicx}

\usepackage{graphics}
\usepackage{amssymb}
\usepackage{amsbsy}
\usepackage{amsfonts}
\usepackage{bm}

\usepackage{amsmath}

\newcommand{\bhat}{\boldsymbol{\hat{b}}}
\newcommand{\bu}{\boldsymbol{u}}

\newcommand{\bb}{\boldsymbol{B}}

\newcommand{\bp}{\boldsymbol{p}}

\newcommand{\pr}{\partial}
\newcommand{\nn}{\nonumber}

\newcommand{\bpar}{\beta_\parallel}
\newcommand{\bee}{\begin{eqnarray}}
\newcommand{\eee}{\end{eqnarray}}

\begin{document}

\title{On the parallel and oblique firehose instability in fluid models.} 

\author{P. Hunana\altaffilmark{1,2} and G. P. Zank\altaffilmark{1,2}}
\altaffiltext{1}{Center for Space Plasma and Aeronomic Research (CSPAR),
University of Alabama, Huntsville, AL 35805, USA}
\altaffiltext{2}{Department of Space Science, University of Alabama, Huntsville, AL 35899, USA}



\begin{abstract}
  A brief analysis of the proton parallel and oblique firehose instability is presented from a fluid perspective and the results are compared
  to kinetic theory solutions obtained by the WHAMP code. It is shown that the classical CGL model very accurately describes the
  growth rate of these instabilities at sufficiently long spatial scales (small wavenumbers). The required stabilization of these instabilities
  at small spatial scales (high wavenumbers) naturally requires dispersive effects and the stabilization is due to the Hall
  term and finite Larmor radius (FLR) corrections to the pressure tensor.
  Even though the stabilization is not completely accurate since at small spatial scales a relatively strong collisionless damping comes into effect,
  we find that the main concepts of the maximum growth rate and the stabilization of these instabilities
  is indeed present in the fluid description. However, there are differences that are quite pronounced when close to the firehose threshold and
  that clarify the different profiles for marginally stable states with a prescribed maximum growth rate $\gamma_{max}$ in the simple fluid models
  considered here and the kinetic description.
\end{abstract}

\maketitle
\section{Introduction}
The importance and utility of the CGL fluid model derived by Chew Goldberger and Low \cite{Chew1956}
that represents the simplest but nevertheless rigorous generalization of
  magnetohydrodynamics (MHD) for collisionless plasmas, is often underestimated and misunderstood by the heliospheric community in general
  and by the solar wind community in particular. Based on an over-simplified analysis of only the mean quantities (where the dispersion relations
  for the fluctuations are disregarded), it is often erroneously claimed that the CGL model predicts an extremely large proton temperature anisotropy far from unity,
  contradicting observational data showing that the solar wind plasma is bounded by the kinetic firehose and mirror instabilities.
  It is then typically concluded that fluid models, such as the CGL model and its generalizations are not particularly useful in studying plasmas with
  anisotropic temperatures. This erroneous claim - that the CGL model predicts extremely large temperature anisotropies - is so omni-present in the solar wind literature,
  that we are inclined to call it the myth of the CGL description. The motivation for this paper is to dispel this myth and to show briefly that the simple CGL
  fluid model is actually a very useful first-order description for plasmas with anisotropic temperatures such as the solar wind,
  since the model captures many first-order anisotropic effects and in fact does not allow extremely large temperature anisotropies. 
  
  Here we concentrate only on the parallel and oblique firehose instability, which are two of the four basic instabilities typically considered in solar wind
  observational studies and in numerous theoretical developments and numerical simulations, the other two being the mirror instability and the ion-cyclotron
  anisotropy instability
  (\cite{Gary1993,Yoon1993,Gary1998,HellingerMatsumoto2000,HollwegIsenberg2002,Kasper2002,WangHau2003,Marsch2004,Marsch2006,Hellinger2006,Matteini2006,Matteini2007,PassotSulem2006,PassotSulem2007,HellingerTravnicek2008,Bale2009,Schekochihin2010,WangHau2010,Rosin2011,Laveder2011,Isenberg2012,PSH2012,SeoughYoon2012,Kunz2014,Hellinger2015,KleinHowes2015,SulemPassot2015,Yoon2015,Matteini2016sw14,Melville2016,Hellinger2017}, and references therein).
  That the CGL model contains the correct parallel and oblique firehose instability threshold is already known from the work of \cite{Abraham-Shrauner1967}
  and the CGL dispersion relations were further studied for example by \citep{FerriereAndre2002,Hunana2013,Hunana2016sw14}.
  The mirror instability and the ion-cyclotron anisotropy instability, are not considered here.
  From a fluid perspective, a lot of attention has been paid especially to the mirror instability, where the CGL description is known to have a factor of
  6 error in the temperature threshold for high plasma beta values. The mirror instability is typically a highly oblique instability and
  the modern Landau fluid description (\cite{PassotSulem2007,PSH2012,SulemPassot2015}, and references therein) that incorporates Landau damping and sophisticated
  finite Larmor radius (FLR) corrections
  captures the entire mirror instability growth rate with excellent accuracy. Considering only the mirror threshold (and not the full wavenumber dependent growth rate),
  the erroneous mirror threshold in the CGL model can also be corrected by simpler fluid models that contain the fluctuating heat flux equations and 
  do not contain Landau damping \citep{PassotRubanSulem2006,DzhalilovKuznetsov2011,Hunana2016sw14}. The parallel and oblique firehose instabilities, even though
  present in the above cited modern Landau fluid description, are usually addressed only very briefly.
  For quasi-parallel propagation directions the higher-order ion-cyclotron damping (in contrast to first order Landau damping) becomes an important
  collisionless damping mechanism and even sophisticated Landau fluid models that are very precise for a more oblique propagation angles, can describe the
  the growth rate of the firehose instability only approximately.
  Here we consider only very simple fluid models based on the CGL description without any form of collisionless damping, the only extensions are the Hall-term
  and the FLR corrections, both of which allow us to explore the firehose instability in sufficient detail.
  The firehose instability in a fluid formalism was also investigated by \cite{WangHau2003,Schekochihin2010,WangHau2010,Rosin2011} and references therein.



\section{The Hall-CGL model with cold electrons}
The nonlinear Hall-CGL model that we use here reads
\begin{eqnarray}
&& \frac{\pr \rho}{\pr t} + \nabla\cdot (\rho \bu )=0; \label{eq:CGL_first}\\
&& \frac{\pr \bu}{\pr t} +\bu \cdot\nabla \bu +\frac{1}{\rho}\nabla\cdot \bp
-\frac{1}{4\pi\rho} (\nabla\times\bb)\times\bb=0;\\
&& \frac{\pr p_\parallel}{\pr t} + \nabla\cdot(p_\parallel \bu) +2p_\parallel \bhat\cdot\nabla\bu\cdot\bhat= 0; \label{eq:PparCGL_cl}\\
&&  \frac{\pr p_\perp}{\pr t} + \nabla\cdot(p_\perp\bu) + p_\perp\nabla\cdot\bu -p_\perp \bhat\cdot\nabla\bu\cdot\bhat=0; \label{eq:PperpCGL_cl}\\
&& \frac{\pr \bb}{\pr t} = \nabla\times(\bu\times\bb)-\frac{c}{4\pi e}\nabla\times \Big(\frac{1}{n}(\nabla\times\bb)\times\bb \Big), \label{eq:CGL_b}
\end{eqnarray}
where the proton pressure tensor $\bp=p_\parallel\bhat\bhat+p_\perp(\boldsymbol{I}-\bhat\bhat)+\boldsymbol{\Pi}$. Since we consider that only protons
contribute to the pressure, we dropped the usual proton index p. The $\bhat=\bb/|\bb|$ is the unit
vector in the direction of the local magnetic field and the $\boldsymbol{\Pi}$ is the finite Larmor radius (FLR) pressure tensor.
If the FLR tensor is neglected and only the first term on the right hand side of the induction equation (\ref{eq:CGL_b}) is considered,
the classical CGL model is recovered. Like MHD, CGL is length-scale invariant.
The second term on the right hand side of the induction equation represents the Hall term and introduces dispersive effects.
The dispersive effects are further refined by the FLR pressure tensor. 
The model is based on numerous approximations. It is assumed that the plasma consists of only protons and electrons,
that it is charge neutral, that the displacement current is neglected, the electrons are cold,
electron inertia is neglected, and the proton heat flux is neglected. The FLR stress forces that enter the pressure equations 
nonlinearly and that are responsible for stochastic heating are neglected. Nevertheless, this very simplified fluid model is completely
sufficient to demonstrate our point.  With the Hall-term present in the induction equation, the pressure equations
(\ref{eq:PparCGL_cl}), (\ref{eq:PperpCGL_cl}) can be rewritten as
\begin{eqnarray}
  \frac{d}{dt}\left( \frac{p_{\parallel} |\bb|^2}{\rho^3}\right) &=&
  \frac{p_{\parallel} |\bb|^2}{\rho^3} \bigg\{-2 \frac{\bhat}{|\bb|}\cdot
    \bigg(\frac{c}{4\pi e}\nabla\times \Big(\frac{1}{n}(\nabla\times\bb)\times\bb\Big)  \bigg)\bigg\}; \label{eq:CGL_c1}\\
  \frac{d}{d t} \left(\frac{p_{\perp}}{\rho |\bb|}\right) &=&  \frac{p_{\perp}}{\rho|\bb|}\bigg\{\frac{\bhat}{|\bb|}\cdot
 \bigg(\frac{c}{4\pi e}\nabla\times \Big(\frac{1}{n}(\nabla\times\bb)\times\bb\Big)  \bigg)
    \bigg\},\label{eq:CGL_c2}
\end{eqnarray}
where $d/dt=\pr/\pr t+\bu\cdot\nabla$ is the convective derivative. If the Hall term is absent in the induction equation, the
right hand sides of (\ref{eq:CGL_c1}) and (\ref{eq:CGL_c2}) are zero, which corresponds to the most often cited form of the CGL pressure equations
with its representation of the adiabatic invariants.
Obviously, the pressure equations (\ref{eq:PparCGL_cl}), (\ref{eq:PperpCGL_cl}) are preferred, since these do not depend on the induction equation and
are valid regardless of whether the Hall term is present or not. Moreover, (\ref{eq:PparCGL_cl}) and (\ref{eq:PperpCGL_cl})
are much easier to linearize and to work with in general.
It is useful to define the usual Alfv\'en speed $V_A=B_0/(4\pi\rho_0)^{1/2}$, the parallel and perpendicular proton plasma beta
and the proton temperature anisotropy ratio $a_p$ as
\begin{equation}
a_p = \frac{T_\perp^{(0)}}{T_\parallel^{(0)}}=\frac{p_\perp^{(0)}}{p_\parallel^{(0)}};\qquad  
\bpar = \frac{v_{th\parallel}^2}{V_A^2}; \qquad \beta_\perp = \frac{v_{th\perp}^2}{V_A^2} = \bpar \frac{T_\perp^{(0)}}{T_\parallel^{(0)}};
\qquad v_{th\parallel\perp} = \sqrt{\frac{2 T_{\parallel\perp}^{(0)}}{m_p}},
\end{equation}
where $ v_{th\parallel\perp}$ are the parallel and perpendicular thermal speeds (importantly, containing the factor of 2). The $(0)$ superscript is used
to emphasize the mean quantities (i.e. to distinguish between the mean pressure $p_{\parallel}^{(0)}$ and the fluctuating pressure $p_\parallel$).
We use the usual convention that the Boltzmann constant $k_B=1$ and
define the temperature as $T_{\parallel\perp}=p_{\parallel\perp}/n$. At long wavelengths ($k V_A/\Omega_p\ll 1$, where $\Omega_p=e B_0/(m_p c)$),
i.e. by neglecting the Hall-term and the FLR corrections, the linearized ($e^{i(\boldsymbol{k}\cdot\boldsymbol{x}-\omega t)}$) CGL model yields forward
and backward propagating oblique Alfv\'en modes satisfying the dispersion relation
\begin{equation} \label{eq:ALF}
\omega = \pm k_\parallel V_A \sqrt{1+\frac{\bpar}{2}(a_p-1)}.
\end{equation}
When the expression under the square root becomes negative the \emph{oblique firehose instability} is obtained and the threshold can be rewritten in many forms,
for example 
\begin{equation} \label{eq:firehose}
1-\frac{\bpar}{2}+\frac{1}{2}\bpar a_p <0; \qquad  \frac{T_\perp^{(0)}}{T_\parallel^{(0)}} + \frac{2}{\bpar}-1 <0; \qquad
\bpar -\beta_\perp >2; \qquad p_\parallel^{(0)}-p_\perp^{(0)} > \frac{B_0^2}{4\pi}.
\end{equation}
The necessary (but not sufficient) condition for the instability to develop is $a_p=T_\perp^{(0)}/T_\parallel^{(0)}<1$.
The first expression directly implies that if $\bpar < 2$, no firehose instability can exist (at these long spatial scales). The firehose instability therefore exists
only for relatively high plasma beta values $\bpar>2$ and only if the background parallel temperature $T_\parallel^{(0)}$ is higher than the perpendicular
temperature $T_\perp^{(0)}$.
The \emph{parallel firehose instability} is the instability of the quasi-parallel whistler mode, and in the parallel propagation limit the threshold is
identical to (\ref{eq:firehose}). For simplicity, here we first consider only the parallel firehose instability for strictly parallel propagation since
the quasi-parallel case requires a longer discussion and will be discussed later in the text. 
The ``hard'' firehose instability threshold (\ref{eq:firehose}) obtained from the CGL model is equivalent to the one found from linear kinetic theory,
which is well known (\cite{Abraham-Shrauner1967,FerriereAndre2002} and references therein).
The parallel firehose instability is usually a propagating instability (i.e. has a nonzero
real frequency) and the oblique firehose instability is usually a non-propagating instability.
In the following figures, where we compare the solutions of linear kinetic theory obtained with the WHAMP code
and the solutions of the Hall-CGL fluid model, we arbitrarily choose $\bpar=4$. This choice is not very convenient since for such a
high plasma beta a strong ion-cyclotron damping is found (which is absent in the fluid models that we discuss) which leads to   
a less precise match between fluid and kinetic results. However, we want to demonstrate our point without resorting to specially tuned cases.
The choice of $\bpar=4$ is convenient in that the firehose threshold is nicely defined with $a_p=T_\perp^{(0)}/T_\parallel^{(0)}<1/2$.
\section{Parallel propagation.}
Without loss of generality, we consider propagation in the x-z plane, where the mean magnetic field $B_0$ is in the z-direction.
Propagation in the x-z plane prescribes that in all expressions $\pr_y=0$.
Here we additionally consider strictly parallel propagation for which $\pr_x=0$. For parallel propagation the sound/ion-acoustic mode decouples
from the rest of the system with the dispersion relation $\omega^2=C_\parallel^2 k_\parallel^2$, and the parallel sound speed
$C_\parallel=\sqrt{3p_\parallel^{(0)}/\rho_0}=V_A\sqrt{3\bpar/2}$.
The dispersion of this mode does not depend on the Hall-term. The dynamics of the other two modes, the ion-cyclotron and the whistler mode, are coupled.
It is useful to define the long-wavelength (CGL) phase speed of the parallel Alfv\'en mode as
\begin{equation}
v_{A\parallel}^2=V_A^2 -\frac{p_\parallel^{(0)}}{\rho_0}+\frac{p_\perp^{(0)}}{\rho_0} = V_A^2 \Big( 1+\frac{\bpar}{2}(a_p-1)\Big).
\end{equation}
For isotropic temperatures ($a_p=1$), $v_{A\parallel}^2=V_A^2$. Linearization of the Hall-CGL system in the x-z plane and
transformation to Fourier space shows that for the parallel propagation
\begin{eqnarray}
\left( \begin{array}{cc}
  \omega^2-v_{A\parallel}^2 k_\parallel^2; &\qquad +i\frac{V_A^2}{\Omega_p}k_\parallel^2 \omega\\
    -i\frac{V_A^2}{\Omega_p}k_\parallel^2 \omega; & \qquad \omega^2-v_{A\parallel}^2 k_\parallel^2
\end{array} \right)
\left( \begin{array}{c} B_x \\ B_y \end{array} \right) = \left( \begin{array}{c} 0 \\ 0 \end{array} \right),
\end{eqnarray}
and the zero determinant requirement can be conveniently rearranged as
\begin{equation}
  \Big( \omega^2+\frac{k_\parallel^2 V_A^2}{\Omega_p}\omega -k_\parallel^2 v_{A\parallel}^2\Big)
  \Big( \omega^2-\frac{k_\parallel^2 V_A^2}{\Omega_p}\omega -k_\parallel^2 v_{A\parallel}^2\Big) = 0. \label{eq:Hall-parF}
\end{equation}
Of the four solutions, two are
\begin{equation}
\omega = \pm \frac{k_\parallel^2 V_A^2}{2\Omega_p} + k_\parallel V_A \sqrt{1+\frac{\bpar}{2}(a_p-1) +\frac{k_\parallel^2V_A^2}{4\Omega_p^2} },\label{eq:Hall-w1}\\
\end{equation}
and the other two are obtained by substituting $\omega$ with $-\omega$ in the above expression. Here we consider only solutions
with positive wavenumbers that yield positive frequencies, and which two solutions to pick depends on if the system is in the firehose-stable or in the
firehose-unstable region. If the quantity $1+\frac{\bpar}{2}(a_p-1)>0$, i.e. if the system is in a firehose-stable regime,
the above solution with the plus sign is the whistler mode (depicted by the blue dashed line in figures that follow)
and the solution with the minus sign is the ion-cyclotron mode (red dashed
line in the following figures). If the quantity $1+\frac{\bpar}{2}(a_p-1)<0$, i.e. if the system is in a firehose-unstable regime, the dispersion relation
(\ref{eq:Hall-w1}) for the whistler
mode is still correct, but the ion-cyclotron mode solution (\ref{eq:Hall-w1}) has to be changed to the mode that was previously backward propagating
with $-\omega$, i.e., solutions in the firehose-unstable region satisfy
\begin{equation}
\omega = \frac{k_\parallel^2 V_A^2}{2\Omega_p} \pm k_\parallel V_A \sqrt{1+\frac{\bpar}{2}(a_p-1) +\frac{k_\parallel^2V_A^2}{4\Omega_p^2} },\label{eq:Hall-wU}\\
\end{equation}
where the solution with the plus sign is the whistler mode (blue) and the solution with the minus sign is the ion-cyclotron mode (red).
For isotropic temperatures $a_p=1$, the dispersion relations do not depend on the value of $\bpar$, which is a consequence of neglecting the FLR pressure corrections, 
as we discuss later. The second term in the dispersion relation (\ref{eq:Hall-wU}) is a purely complex number in the region where 
\begin{equation} \label{eq:HallCGL-firehose}
1+\frac{\bpar}{2}(a_p-1) +\frac{k_\parallel^2V_A^2}{4\Omega_p^2} <0,
\end{equation}
which can be regarded as a modified Hall-CGL firehose instability threshold that is now length-scale dependent. At long spatial scales ($k_\parallel\rightarrow 0$),
the above criterion is simply the CGL firehose condition. In the region where (\ref{eq:HallCGL-firehose}) is satisfied,
both modes (\ref{eq:Hall-wU}) propagate with the same real frequency, the first term $\omega_r=k_\parallel^2 V_A^2/(2\Omega_p)$, and both modes have
the same absolute value of the imaginary frequency, the second term
$\omega_i=\pm k_\parallel V_A \sqrt{-1-\frac{\bpar}{2}(a_p-1) -\frac{k_\parallel^2V_A^2}{4\Omega_p^2} }$, and the total frequency $\omega=\omega_r+i\omega_i$.
The important difference is that the whistler mode has a positive growth rate and is unstable, and the ion-cyclotron mode has negative growth rate and is stable and damped.
The unstable whistler mode represents the parallel firehose instability. 
Above some critical wavenumber the left hand side of (\ref{eq:HallCGL-firehose}) will become positive and the frequency of both modes is then
purely real. The Hall-term therefore stabilizes the firehose instability at sufficiently high wavenumbers (small spatial scales). 

The results of fluid models are often compared to those from a kinetic description. In discussing ``resonances'', fluid models are typically formulated 
in a way that singularities are ``removed'' from the dispersion relations (usually as a consequence of eliminating the electric field when calculating
the dispersion relations, although fluid models can be naturally derived also through the electric field eigenmatrix, see for example \cite{Sahraoui2012,Zank2014}).
The dispersion equation (\ref{eq:Hall-parF}) for parallel propagation is easily expressed as
\begin{equation}
  \frac{k_\parallel^2}{\omega^2} = -\frac{\Omega_p/V_A^2}{\omega- \Omega_p\big(1+\frac{\bpar}{2}(a_p-1)) }; \qquad
  \frac{k_\parallel^2}{\omega^2} = +\frac{\Omega_p/V_A^2}{\omega+ \Omega_p\big(1+\frac{\bpar}{2}(a_p-1)) },
\end{equation}
which for the isotropic temperatures ($a_p=1$) or in the cold plasma limit ($\bpar=0$) simplifies to
\begin{equation}
  \frac{k_\parallel^2}{\omega^2} = -\frac{\Omega_p/V_A^2}{\omega- \Omega_p}; \qquad
  \frac{k_\parallel^2}{\omega^2} = +\frac{\Omega_p/V_A^2}{\omega+ \Omega_p},
\end{equation}
making it obvious that the simplest ion-cyclotron resonances are present in the Hall-CGL description as well. 

To better understand what effects are present and what effects are absent in the Hall-CGL model, Figure 1 shows solutions of the full linear kinetic theory
obtained with the WHAMP code (assuming a bi-Maxwellian distribution function).
From now on we drop the superscript $(0)$ for the background temperature. 
The proton temperature is assumed to be isotropic $T_\perp/T_\parallel=1$ and the electrons to be cold with $T_e/T_p=0$ (in the WHAMP code the value is chosen to be
$10^{-8}$). In the WHAMP code it is
necessary to prescribe the ratio of the parallel proton thermal speed to the speed of light (because the displacement current is present) and we choose $v_{th\parallel}/c = 10^{-4}$.
The solid lines are the solutions of the kinetic theory with different values $\bpar=10^{-4}; 0.1; 1; 2; 4$.
The blue curve is the whistler mode and the red curve is the ion-cyclotron
mode. For isotropic temperatures the solutions of the Hall-CGL model are completely $\bpar$ independent and are shown by two dashed curves. The kinetic
solutions for the whistler mode are almost $\bpar$ independent and only the mode with $\bpar=4$ deviates from the other whistler solutions,
and only between $k V_A/\Omega_p=0.1-2.0$.
On the other hand, the ion-cyclotron mode is strongly $\bpar$ dependent, partially because of the FLR corrections and partially because the mode
is strongly ion-cyclotron damped, which is not captured in the fluid models considered here. In the next section, we consider the simple FLR corrections and the
solutions are shown in Figure 1 right; discussed below.
\begin{figure*}
$$\includegraphics[width=0.48\linewidth]{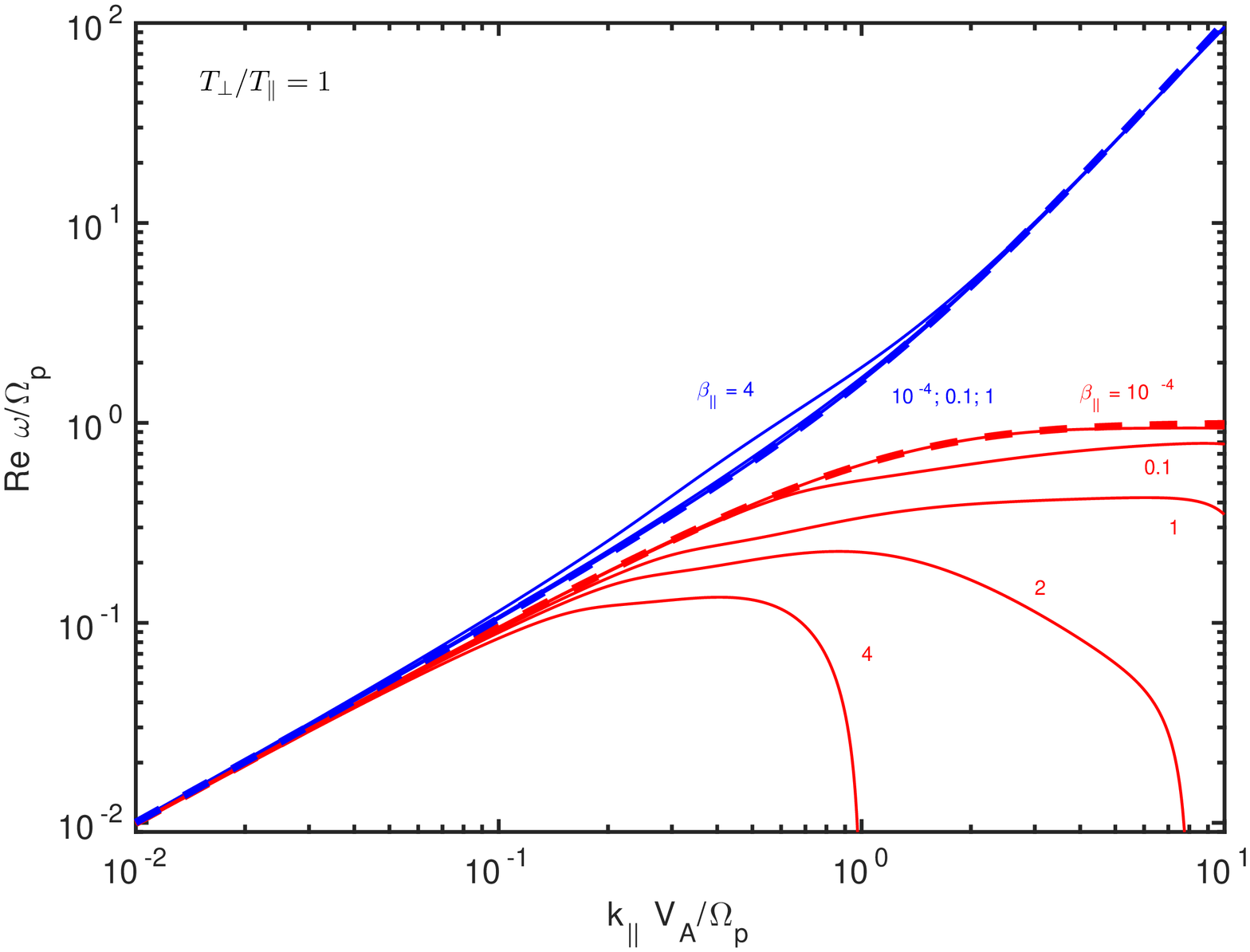}\hspace{0.03\textwidth}\includegraphics[width=0.48\linewidth]{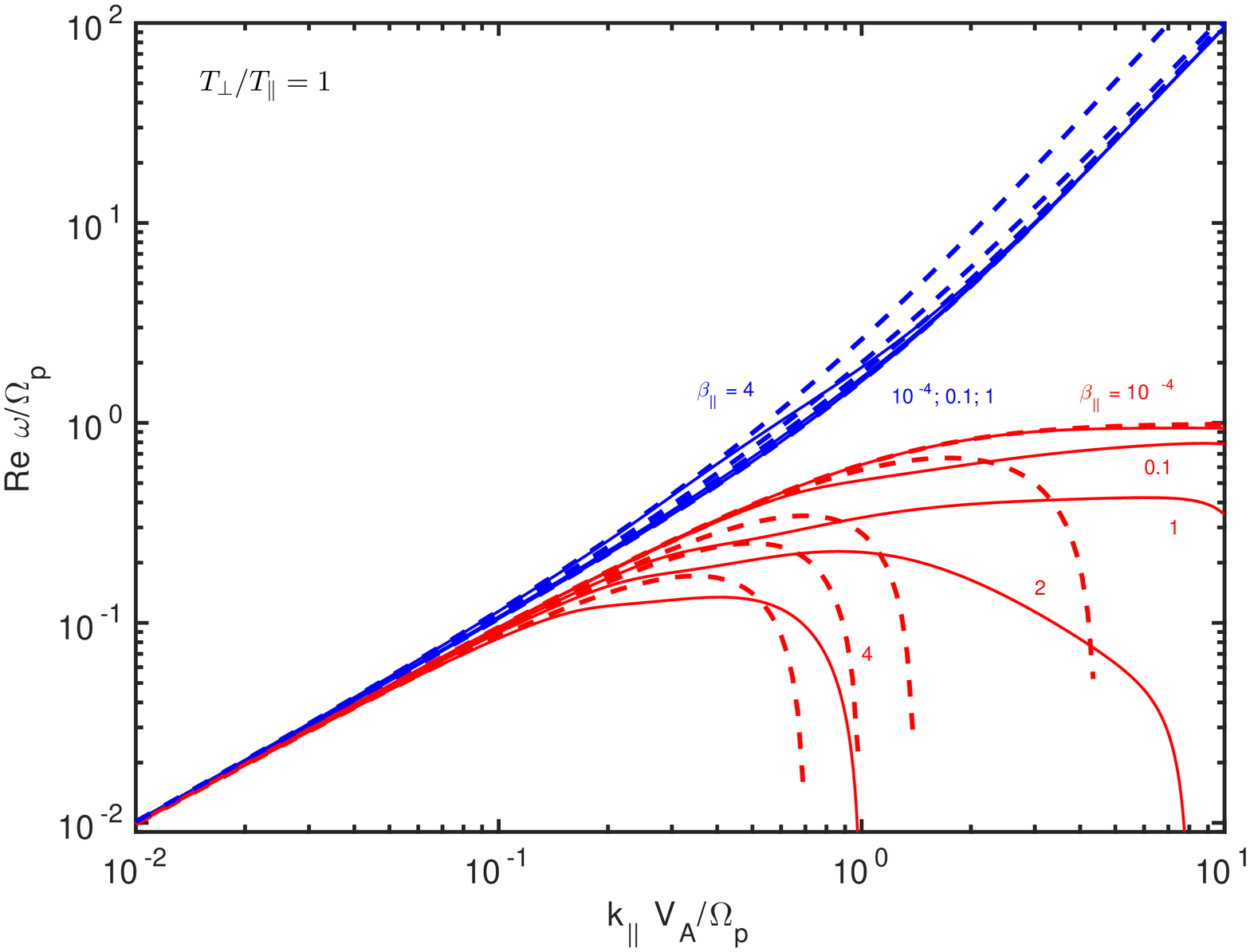}$$
  \caption{Left: Comparison of solutions of the full kinetic theory obtained by the WHAMP code (solid lines) and the dispersion relation of the Hall-CGL model (dashed lines)
    for parallel propagating whistler modes (blue) and ion-cyclotron modes (red). The proton temperature is isotropic $T_\perp/T_\parallel=1$ and the electrons are cold.
    Real frequency is plotted as a function of parallel wavenumber. The $\bpar$ is varied as $\bpar=10^{-4}; 0.1; 1; 2; 4$. The Hall-CGL
    ion-cyclotron mode matches the kinetic ion-cyclotron mode only for the smallest $\bpar=10^{-4}$. Right: The same kinetic solutions as in the left figure are compared to
    the Hall-CGL model with simple FLR corrections. It is obvious that the first order FLR corrections used are ``too strong'' and introduce errors at high wavenumbers and frequencies.
    Nevertheless, it is interesting that the real frequency for the fluid ion-cyclotron resonances is crudely consistent. The second order FLR corrections (not shown) significantly
    eliminate the errors introduced for the whistler mode, but do not improve the ion-cyclotron mode dispersions.} 
\end{figure*}

Figures 2-6 show the parallel firehose instability for a fixed plasma beta $\bpar=4$ and a variable temperature anisotropy $T_\perp/T_\parallel$.
In Figure 2, the temperature anisotropy is varied in a way that the modes remain in a firehose stable regime. 
Evidently, for sufficiently low wavenumbers, the Hall-CGL model accurately reproduces the real frequency
of both modes. For high wavenumbers only the whistler mode is reproduced  reasonably accurately by the Hall-CGL model and the ion-cyclotron mode strongly deviates
from the kinetic results (similarly to Figure 1
the behavior can be partially reproduced by introducing FLR corrections, which we do not show).        
In Figure 3, the temperature anisotropy is varied so that the modes are in a firehose unstable regime and the imaginary phase speed is plotted. The
ion-cyclotron mode is damped and the whistler mode is unstable. At low wavenumbers, the Hall-CGL model reproduces the imaginary phase speed very accurately for both
modes for all the values of $T_\perp/T_\parallel$ considered. At higher wavenumbers the firehose instability of the whistler mode is stabilized and in the Hall-CGL model
this is a consequence of the Hall-term. The stabilization can be improved by considering first-order FLR corrections, which is shown in Figure 4. At higher wavenumbers
the ion-cyclotron mode experiences strong ion-cyclotron damping in the kinetic description, whereas in the fluid models considered here the damping of the
ion-cyclotron mode vanishes at a wavenumber where the whistler firehose instability is stabilized. The parallel firehose instability is usually a propagating
instability and in Figure 5 we plot the real frequency. We plot the dispersion relation for the Hall-CGL fluid model, together with two different fluid models using first-order
FLR and second-order FLR corrections, discussed in the next section. Finally, to clearly see the differences in the maximum growth rate,
in Figure 6 we directly plot the damping rate for the kinetic and Hall-CGL models, and use a linear scale for the wavenumber axis. 

\begin{figure*}
$$\includegraphics[width=0.48\linewidth]{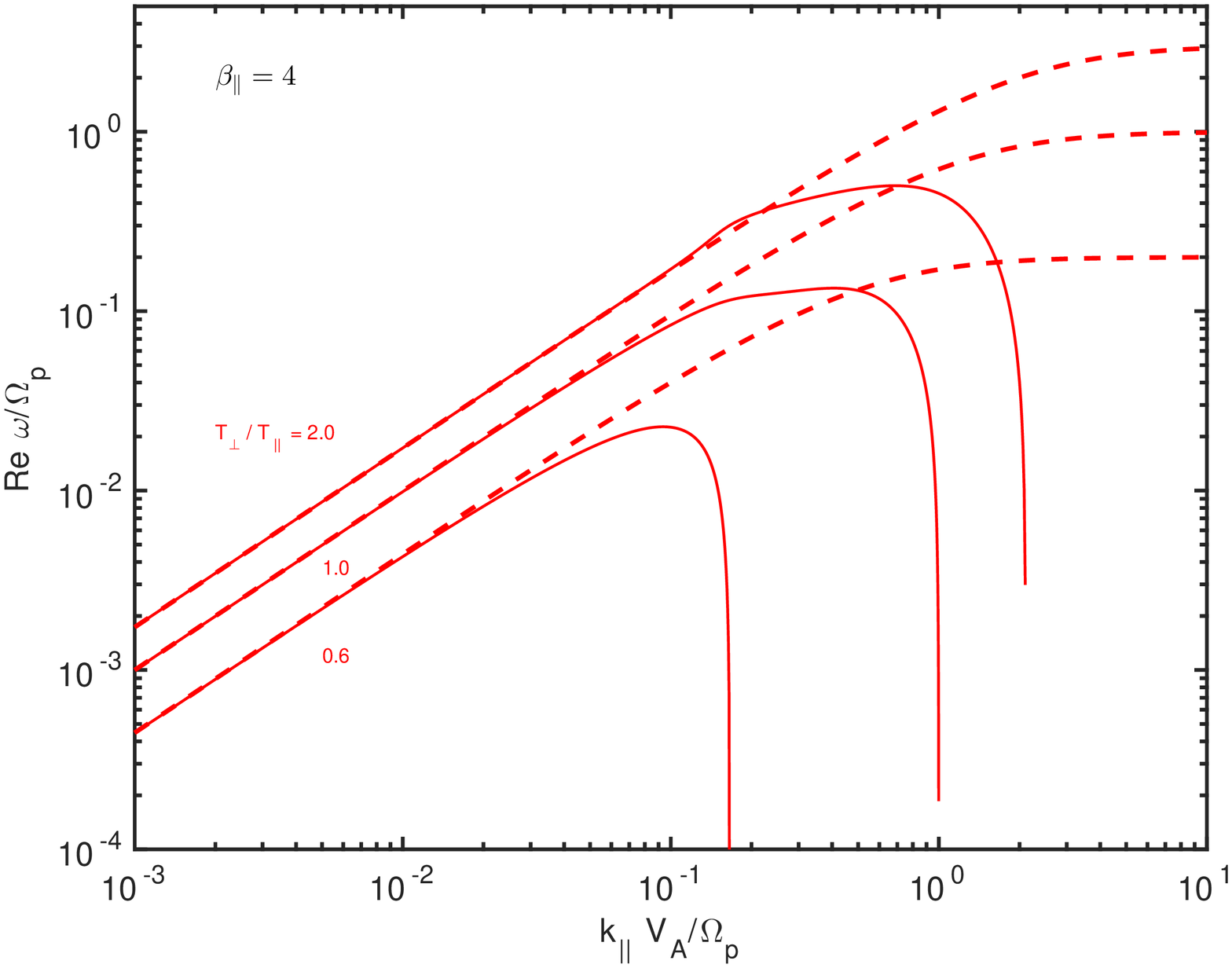}\hspace{0.03\textwidth}\includegraphics[width=0.48\linewidth]{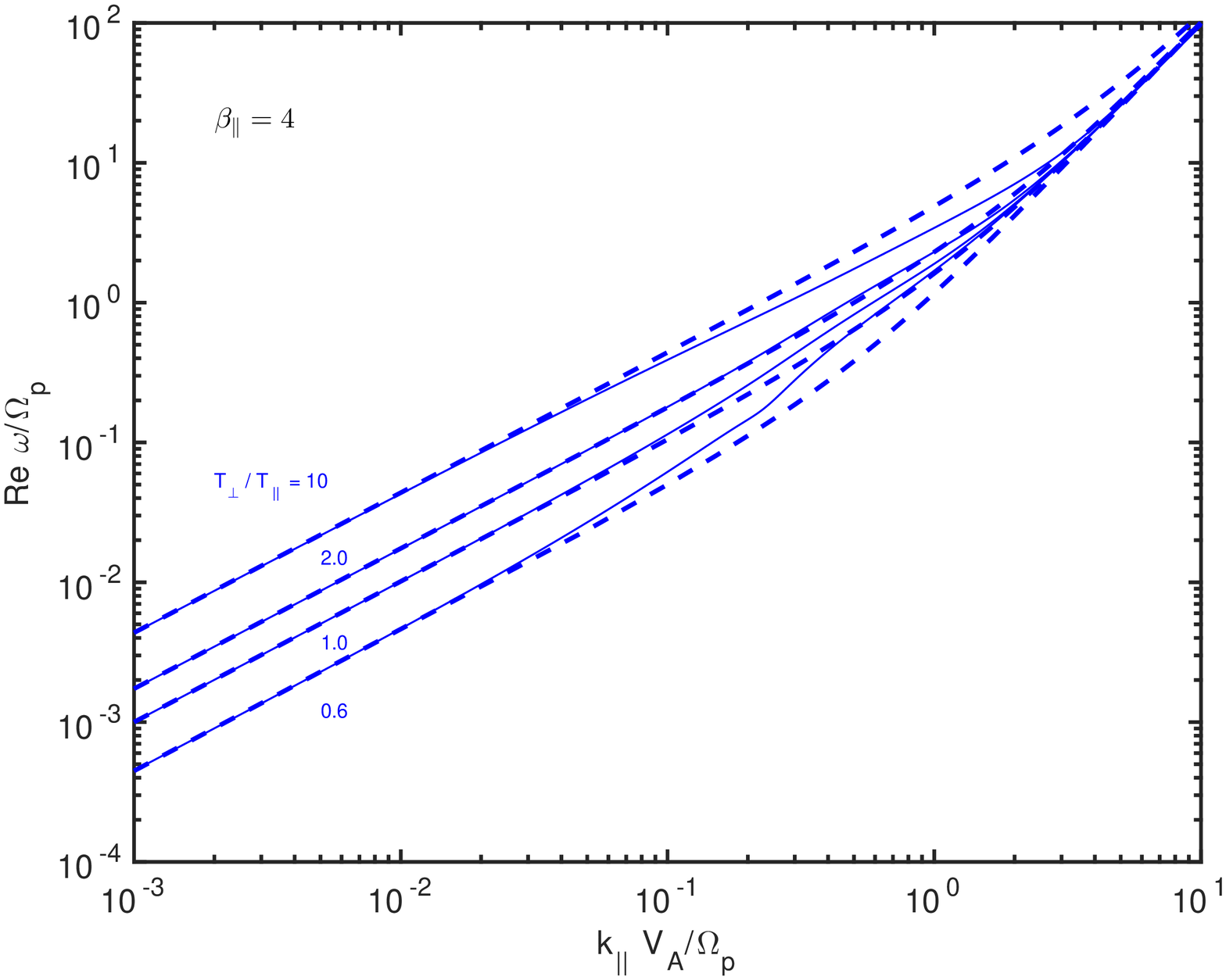}$$
  \caption{Firehose stable regime with $\bpar=4$ for parallel propagation. The ion-cyclotron mode (red) and whistler mode (blue) of kinetic theory obtained from the WHAMP
    code (solid lines) and the dispersion relation of the Hall-CGL model (dashed lines) are shown. Real frequency is plotted.
    The proton temperature anisotropy is varied as $T_\perp/T_\parallel=2.0; 1.0; 0.6$ (and also $T_\perp/T_\parallel=10$ for the whistler mode),
    ensuring that the firehose threshold $T_\perp/T_\parallel=0.5$ is never reached.} 
\end{figure*}

\begin{figure*}
$$\includegraphics[width=0.48\linewidth]{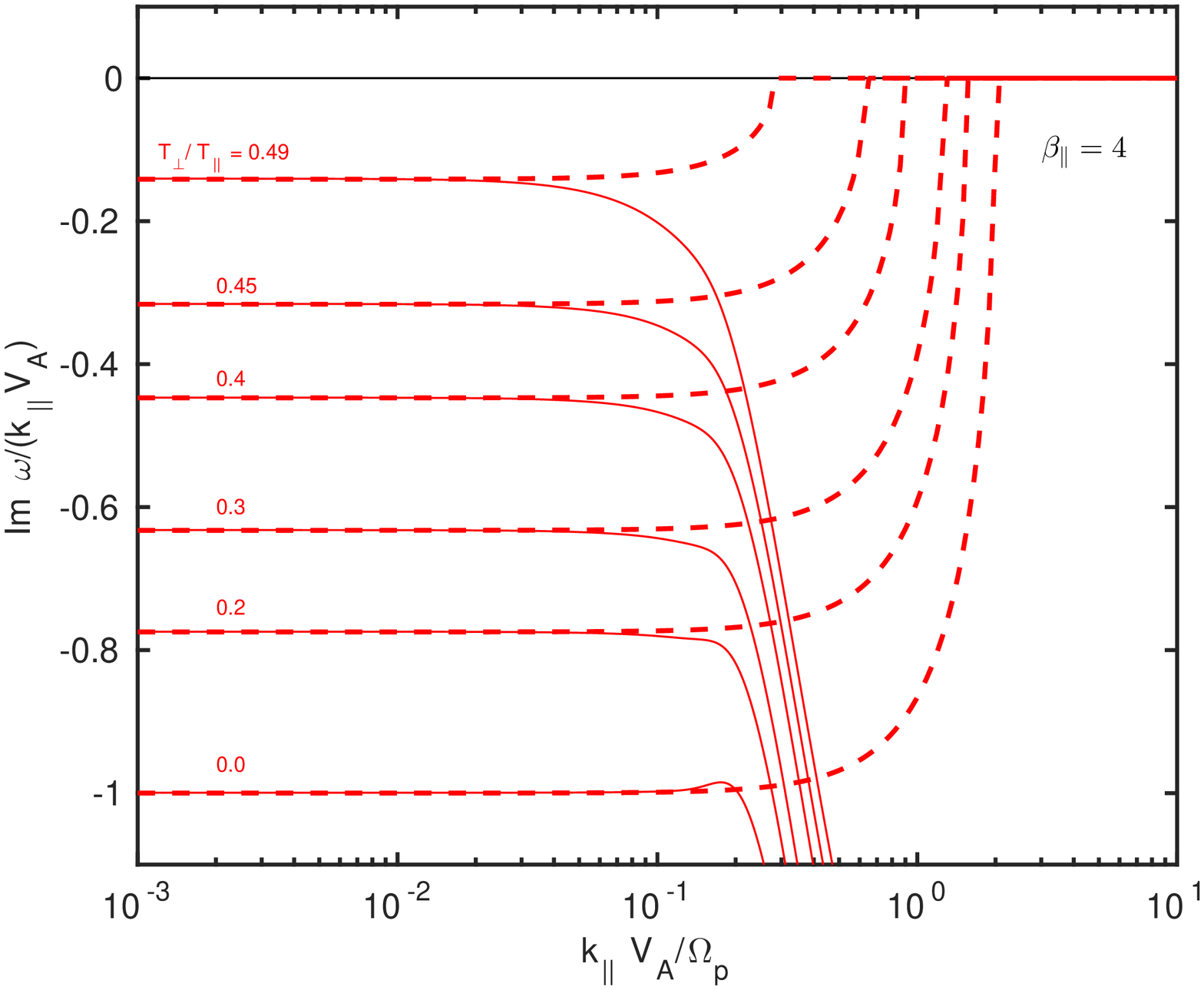}\hspace{0.03\textwidth}\includegraphics[width=0.48\linewidth]{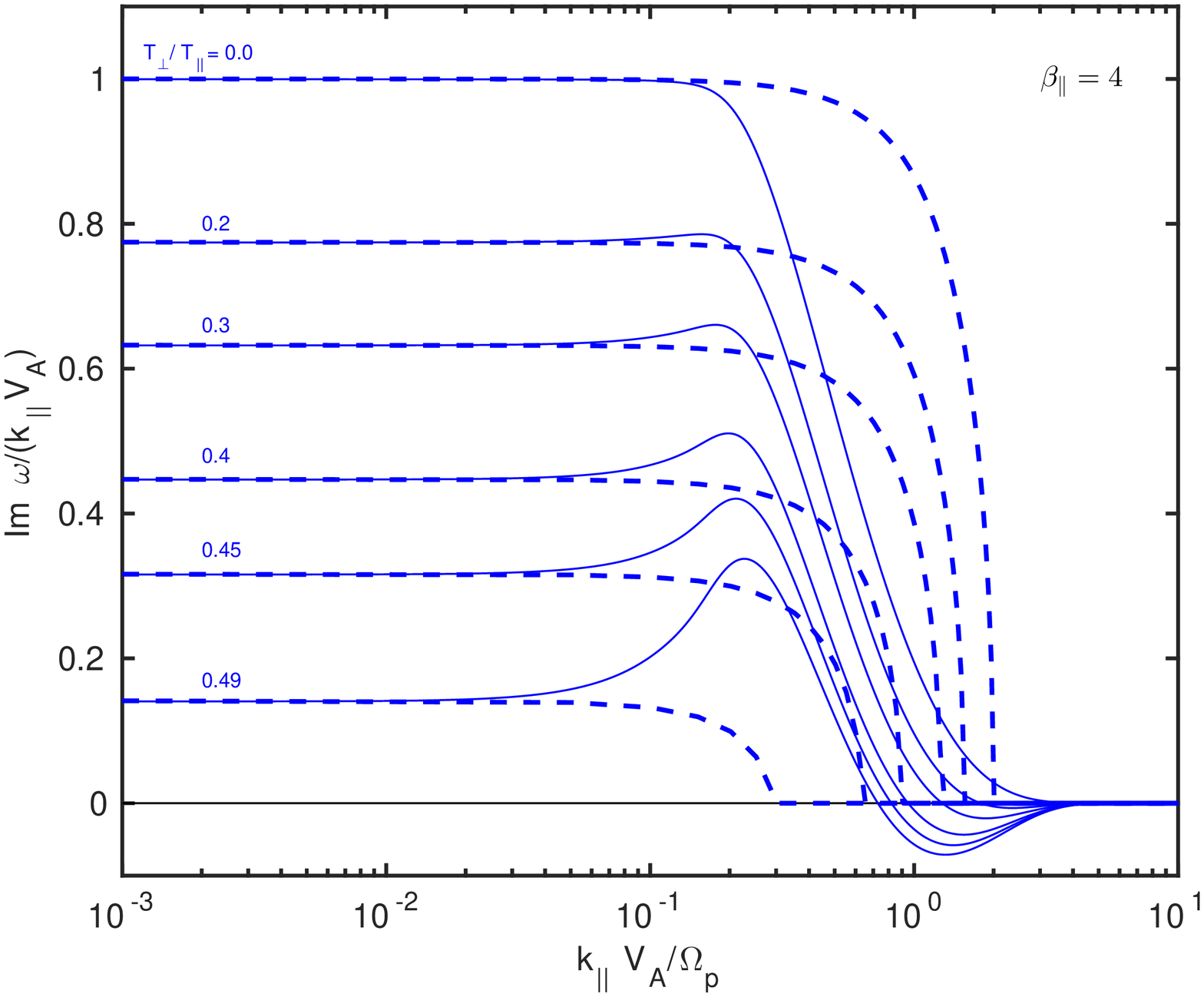}$$
  \caption{Firehose unstable regime with $\bpar=4$ after crossing the threshold $T_\perp/T_\parallel=0.5$. The temperature anisotropy is varied as
    $T_\perp/T_\parallel=0.49; 0.45; 0.4; 0.3; 0.2; 0.0$. The WHAMP code cannot be run with the anisotropy $T_\perp/T_\parallel=0.0$ exactly and a small value of
    $10^{-4}$ is chosen instead. The imaginary frequency normalized to the wavenumber (imaginary phase speed) is plotted.
    Left: the stable ion-cyclotron mode. Right: the unstable whistler mode.
    Solid lines are solutions of kinetic theory and dashed lines are solutions of the Hall-CGL model. At long-wavelengths (small wavenumbers),
    the Hall-CGL model (and the CGL model) accurately describe the growth rate of the parallel firehose instability of the whistler mode.
    It also accurately describes the damping of the ion-cyclotron mode at long-wavelengths, because the damping of the ion-cyclotron mode at these scales comes
    primarily from the coupling to the whistler firehose instability mechanism and not from collisionless ion-cyclotron damping.} 
\end{figure*}

\begin{figure*}
$$\includegraphics[width=0.48\linewidth]{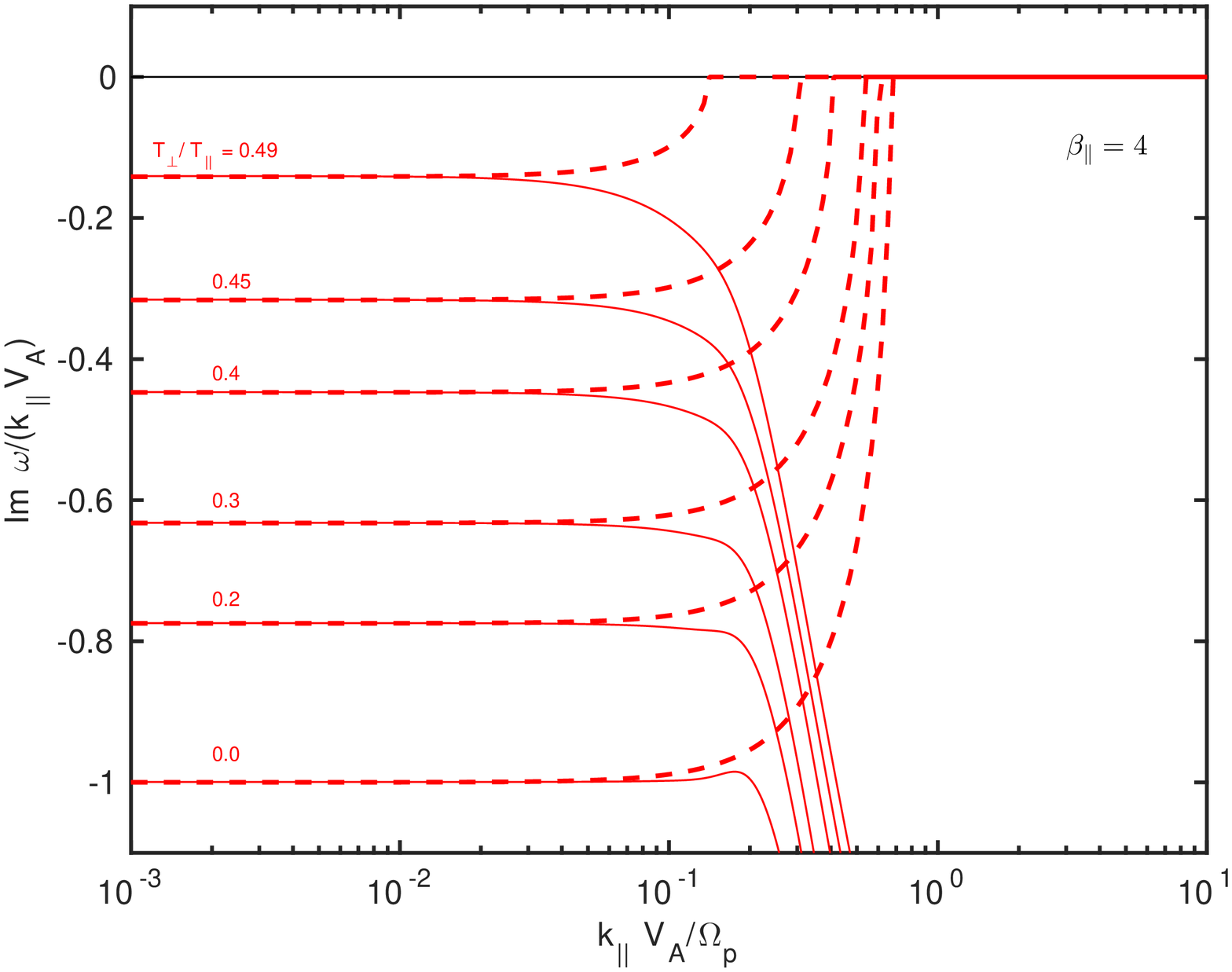}\hspace{0.03\textwidth}\includegraphics[width=0.48\linewidth]{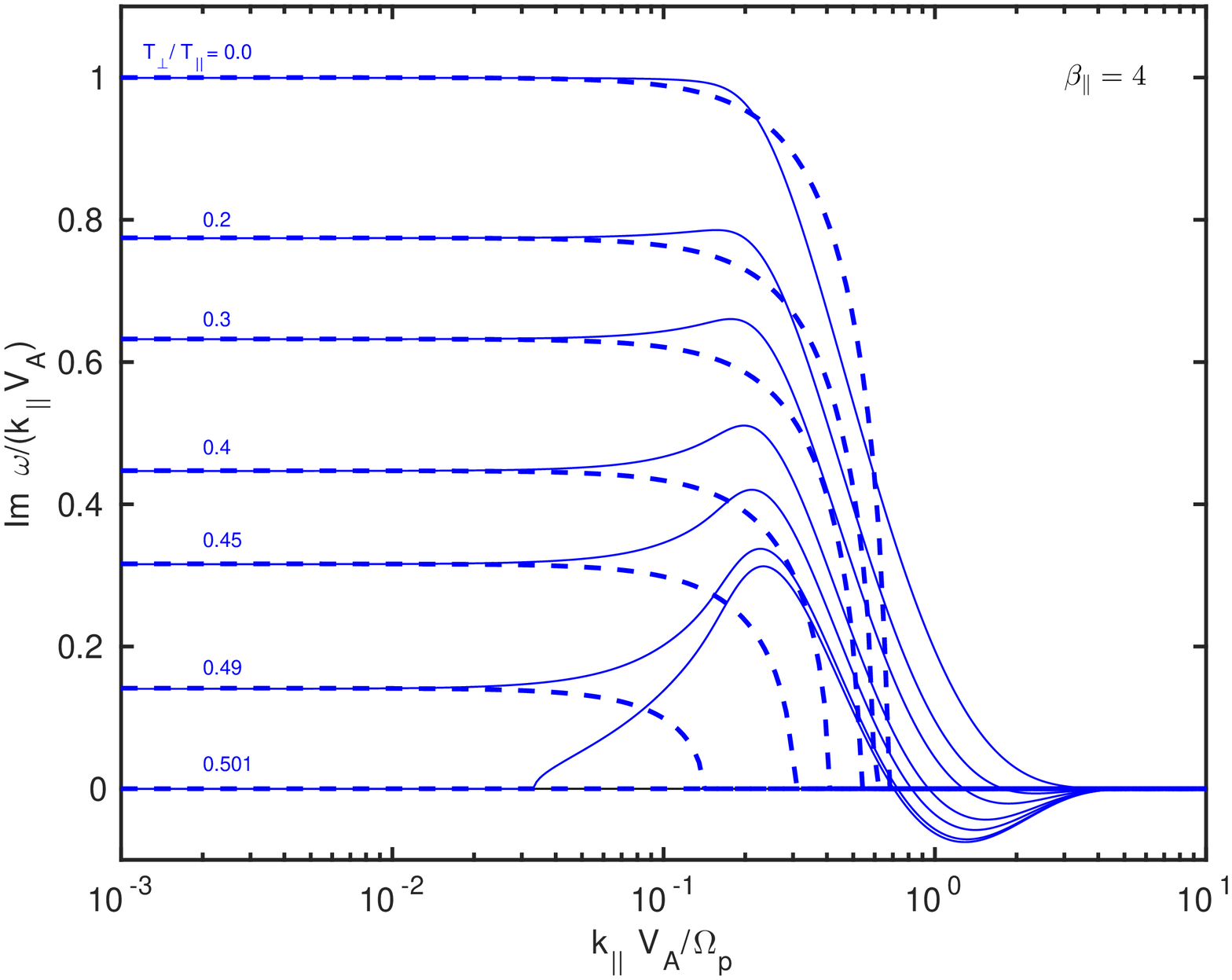}$$
  \caption{Same as Figure 3, except that the Hall-CGL fluid model now includes the simple FLR corrections. It can be seen that the stabilization of the parallel firehose
    instability is slightly improved. In the right figure, a further solution with $T_\perp/T_\parallel=0.501$ is added, which represents a solution in the firehose-stable regime
    (at small wavenumbers). It is seen that for this solution the fluid model again accurately captures the growth rate at small wavenumbers where both the fluid model and the
    kinetic theory have zero growth rate, implying that the ``hard'' firehose threshold (\ref{eq:firehose}) derived for $k\rightarrow 0$ is indeed correct.
    However, the fluid model remains stable for higher wavenumbers and the kinetic solution becomes firehose unstable at some range of wavenumbers. This effect is responsible for
    the differences in ``hard'' firehose thresholds derived for $k\rightarrow 0$ that are correctly described already by the CGL model, and the kinetic contours obtained for
    a prescribed growth rate $\gamma_{max}$.} 
\end{figure*}

\begin{figure*}
$$\includegraphics[width=0.48\linewidth]{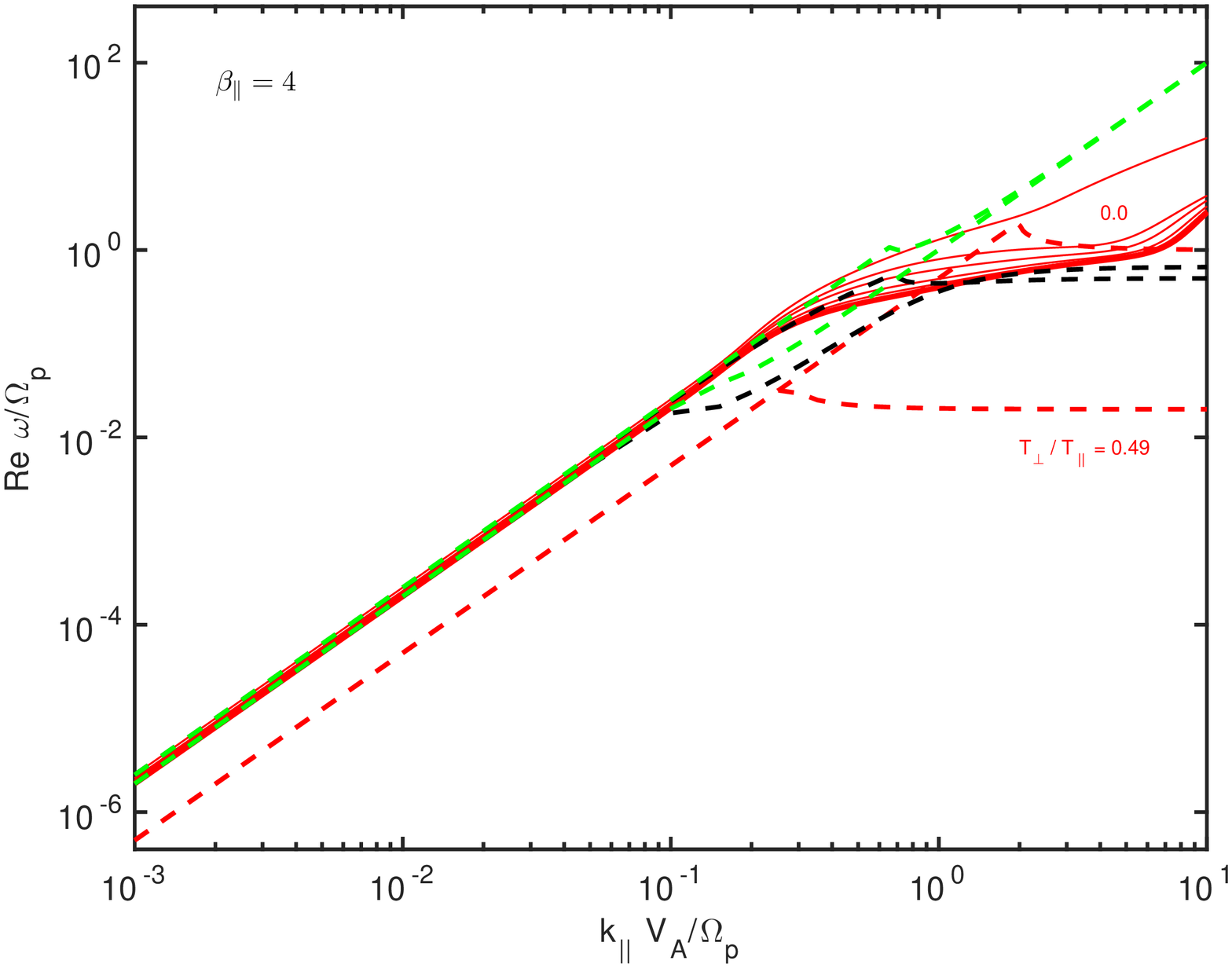}\hspace{0.03\textwidth}\includegraphics[width=0.48\linewidth]{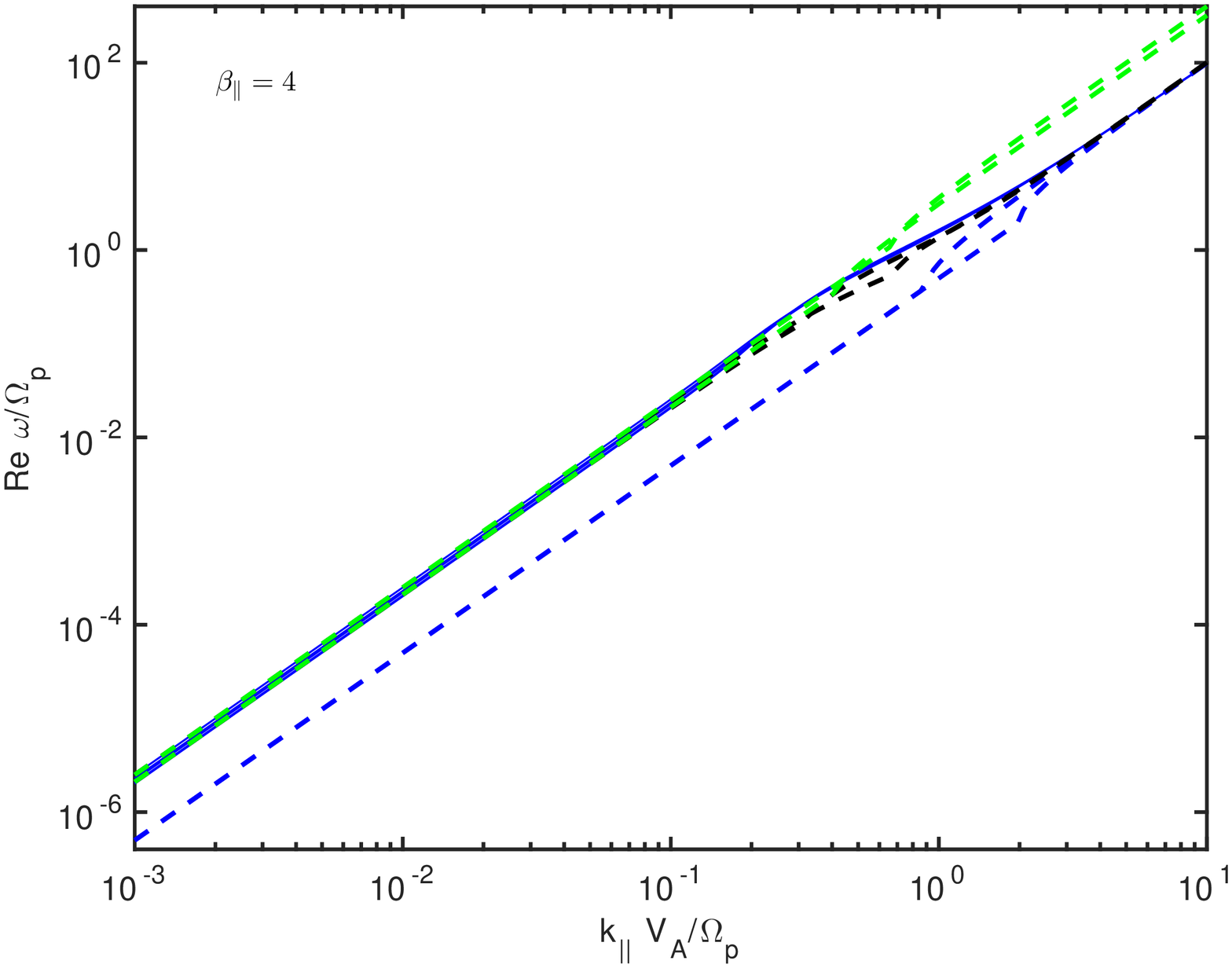}$$
  \caption{Real frequency in the firehose-unstable regime since the parallel firehose instability is usually a propagating instability.
    Left: the ion-cyclotron mode; Right: the whistler mode. Kinetic solutions are shown by solid lines and all solutions with  $T_\perp/T_\parallel=0.49; 0.45; 0.4; 0.3; 0.2; 0.0$
    are plotted. The dashed lines show fluid solutions and to prevent cluttering with too many lines, only two solutions with the most extreme values of
    $T_\perp/T_\parallel=0.49$ and $T_\perp/T_\parallel=0.0$ are plotted for each fluid model. All other solutions lie between these two values. 
    The red dashed lines (left figure) and the blue dashed lines (right figure) correspond to the Hall-CGL model. The figure shows that the Hall-CGL model describes the
    real frequency of both modes in the firehose-unstable regime only approximately, even at long wavelengths, since the frequency is very small.
    The dashed green lines are solutions that include the first order FLR corrections, illustrating that these solutions match the kinetic results accurately at long wavelengths.
    However, at short wavelengths the simple FLR corrections overestimate the real frequency. The dashed black lines represent the second order FLR corrections and
    these solutions improve the accuracy of the solutions at short wavelengths, especially for the whistler mode.} 
\end{figure*}

\section{FLR corrections}
We use the simplest well-known anisotropic FLR tensor evaluated along $\bhat=(0,0,1)$, where
\begin{eqnarray}
&& \Pi_{xx} = -\Pi_{yy} = -\frac{p_\perp^{(0)}}{2\Omega_p} (\pr_x u_y + \pr_y u_x);\nn\\
  && \Pi_{xy} = \frac{p_\perp^{(0)}}{2\Omega_p} (\pr_x u_x-\pr_y u_y);\nn\\
  && \Pi_{xz} = -\frac{1}{\Omega_p}\Big( (2p_\parallel^{(0)}-p_\perp^{(0)})\pr_z u_y + p_\perp^{(0)} \pr_y u_z \Big);\nn\\
  && \Pi_{yz} = \frac{1}{\Omega_p}\Big( (2p_\parallel^{(0)}-p_\perp^{(0)})\pr_z u_x + p_\perp^{(0)} \pr_x u_z \Big);\nn\\
  && \Pi_{zz} =0. \label{eq:FLR1}
\end{eqnarray}
For parallel propagation the form is especially simple since $\Pi_{xx}=\Pi_{xy}=0$ and $\Pi_{xz},\Pi_{yz}$ are simplified.
The parallel whistler and ion-cyclotron modes dispersion relations read
\begin{equation}
  \omega = \pm \frac{k_\parallel^2 V_A^2}{2\Omega_p}\Big(1+\bpar(1-\frac{a_p}{2})\Big) + k_\parallel V_A \sqrt{1+\frac{\bpar}{2}(a_p-1)
    +\frac{k_\parallel^2V_A^2}{4\Omega_p^2} \Big(1-\bpar(1-\frac{a_p}{2}) \Big)^2 },\label{eq:FLR-w1}\\
\end{equation}
and another two solutions are obtained by substituting $\omega$ with $-\omega$. The solution agrees with equation (A1) in \cite{Hunana2011}
and it is also valid for the simplest Landau fluid model with FLR corrections (\ref{eq:FLR1}), since at the linear level,
the strictly parallel propagating whistler and ion-cyclotron modes are not affected by Landau damping.
As before, for the ion-cyclotron mode in the firehose-unstable regime, the $-\omega$ solution has to be chosen. 
These analytic solutions are used in Figure 1 right, Figure 4, Figure 5 and Figure 6 right.
The simple FLR corrections are useful for describing the growth rate more accurately (although not when close to the firehose threshold) and in
improving the accuracy at long wavelengths (low k) in the real frequency of the parallel firehose instability of the whistler mode.
However, the FLR corrections (\ref{eq:FLR1}) are inaccurate for higher frequencies and introduce errors that were not present in the Hall-CGL model, as is clearly visible
when plotting the real frequency of the whistler mode at high wavenumbers. A more accurate representation of the high frequency whistler mode can be obtained by considering
higher-order FLR corrections obtained by expanding the full pressure tensor equation. Following \cite{Goswami2005}, who considered much more general FLR corrections
including the heat flux contributions, which we neglect here (see their eq. 18-21), the components of the FLR tensor satisfy (written in the linear approximation)
\begin{eqnarray}
&&  \frac{\pr \Pi_{xx}}{\pr t} -2\Omega_p\Pi_{xy} + p_\perp^{(0)}(\pr_x u_x-\pr_y u_y) =0;\\
&&  \frac{\pr \Pi_{xy}}{\pr t} +2\Omega_p\Pi_{xx} + p_\perp^{(0)}(\pr_x u_y+\pr_y u_x) =0;\\
&&  \frac{\pr \Pi_{xz}}{\pr t} -\Omega_p\Pi_{yz} + p_\perp^{(0)}\pr_x u_z + p_\parallel^{(0)}\pr_z u_x +(p_{\parallel}^{(0)}-p_\perp^{(0)})\frac{1}{B_0}\frac{\pr B_x}{\pr t} =0;\\
&&  \frac{\pr \Pi_{yz}}{\pr t} +\Omega_p\Pi_{xz} + p_\perp^{(0)}\pr_y u_z + p_\parallel^{(0)}\pr_z u_y +(p_{\parallel}^{(0)}-p_\perp^{(0)})\frac{1}{B_0}\frac{\pr B_y}{\pr t} =0,
\end{eqnarray}
and $\Pi_{yy}=-\Pi_{xx}$; $\Pi_{zz}=0$.
When the time derivatives of the FLR tensor are neglected and only the first term in the linearized induction equation is used (i.e. when the Hall term is neglected),
the first order FLR tensor (\ref{eq:FLR1}) is obtained. To obtain a more precise FLR tensor, the tensor can be decomposed into first and second
order terms $\Pi=\Pi^{(1)}+\Pi^{(2)}$. By neglecting $\pr \Pi^{(2)}/\pr t$ in the above equations, the $\Pi^{(2)}$ components are obtained. One can choose to move the
Hall-term contributions to $\Pi^{(2)}$ so that the $\Pi^{(1)}$ is equivalent to (\ref{eq:FLR1}), or one can keep the Hall contributions in $\Pi^{(1)}$.
At least for parallel propagation and for the range of parameters explored here, the first choice (i.e. when the time derivative of the
Hall-term is neglected) yields better results. Introducing for brevity the normalized wavenumbers and frequencies
$\widetilde{\omega}=\omega/\Omega_p$, $\widetilde{k}=kV_A/\Omega_p$ and the expression $v_b=\bpar(1-a_p/2)$, allows the 
parallel propagating whistler and ion-cyclotron modes to be expressed as
\begin{eqnarray}
  \widetilde{\omega} = \frac{1}{1+v_b \widetilde{k}_\parallel^2} \bigg[ \pm \frac{\widetilde{k}_\parallel^2}{2}\Big( 1+v_b(1+\widetilde{k}_\parallel^2) \Big)
    +\widetilde{k}_\parallel \sqrt{ \Big(1+\frac{\bpar}{2}(a_p-1)-\frac{\bpar}{2}\widetilde{k}_\parallel^2\Big)(1+v_b\widetilde{k}_\parallel^2)+
    \frac{\widetilde{k}_\parallel^2}{4}\Big( 1+v_b(1+\widetilde{k}_\parallel^2)  \Big)^2 } \bigg],
\end{eqnarray}  
with two further solutions corresponding to $-\widetilde{\omega}$ on the left hand side. Again, for the ion-cyclotron mode in the firehose-unstable regime,
the $-\widetilde{\omega}$ solution has to be chosen. These analytic solutions are used only in Figure 5.
For completeness, the second approach (i.e. when the time derivative of the Hall term is not neglected) yields the dispersion relation
\begin{eqnarray}
  \widetilde{\omega} = \frac{1}{1+v_b \widetilde{k}_\parallel^2} \bigg[ \pm \frac{\widetilde{k}_\parallel^2}{2}\Big( 1+v_b+\frac{\bpar}{2}\widetilde{k}_\parallel^2 \Big)
    +\widetilde{k}_\parallel \sqrt{ \Big(1+\frac{\bpar}{2}(a_p-1)-\frac{\bpar}{2}\widetilde{k}_\parallel^2\Big)(1+v_b\widetilde{k}_\parallel^2)+
    \frac{\widetilde{k}_\parallel^2}{4}\Big( 1+v_b+\frac{\bpar}{2}\widetilde{k}_\parallel^2  \Big)^2 } \bigg],
\end{eqnarray}
but we do not plot this solution. Notice that for isotropic temperatures $a_p=1$ both the first and second
derivation of the dispersion relations are equal (since $v_b=\bpar/2$), which is expected since for $p_{\parallel}^{(0)}=p_\perp^{(0)}$ the contributions from the induction equation
entering the FLR corrections vanish.
\section{The maximum growth rate for parallel propagation}
For the whistler mode in the Hall-CGL model it is easy to find the maximum growth rate from $\pr \omega_i/\pr k =0$. It is useful
to work with normalized frequencies and wavenumbers, $\widetilde{\omega}=\omega/\Omega_p$, $\widetilde{k}=kV_A/\Omega_p$. Assuming that the whistler mode
is in the firehose unstable regime, the maximum growth rate occurs at $(\widetilde{k}_\parallel^2)_{max}=-2(1+\frac{\bpar}{2}(a_p-1))$,
and the maximum growth rate of the parallel firehose instability in the Hall-CGL model is 
\begin{equation}
\gamma_{max}\equiv \left(\frac{\omega_i}{\Omega_p}\right)= -\Big(1+\frac{\bpar}{2}(a_p-1)\Big).
\end{equation}
The right hand side is obviously the ``hard'' threshold obtained for the exactly zero growth rate. As the maximum growth rate
increases, the thresholds are naturally modified.
For the Hall-CGL model with the first order FLR corrections, the derivative $\pr \omega_i/\pr k =0$ yields the maximum growth rate at
$(\widetilde{k}_\parallel^2)_{max}=-2(1+\frac{\bpar}{2}(a_p-1))/(1+\frac{\bpar}{2}(a_p-2))^2$, from which the maximum growth rate is 
\begin{equation}
\gamma_{max}=\frac{1+\frac{\bpar}{2}(a_p-1)}{1+\frac{\bpar}{2}(a_p-2)}.
\end{equation}
The growth rates are shown in Figure 6, where it is shown that a maximum growth rate indeed exists in the Hall-CGL model (left figure) -
this is not completely obvious from the previous Figures where the phase speed is plotted. A linear scale for the wavenumber axis is used.
The accuracy of the locations and size of the maximum is improved with the inclusion of the first order FLR corrections (i.e., Hall-CGL-FLR1 model, right figure).
Nevertheless, large deviations from the kinetic model maxima exist, especially for solutions close to the threshold. 
\begin{figure*}
$$\includegraphics[width=0.48\linewidth]{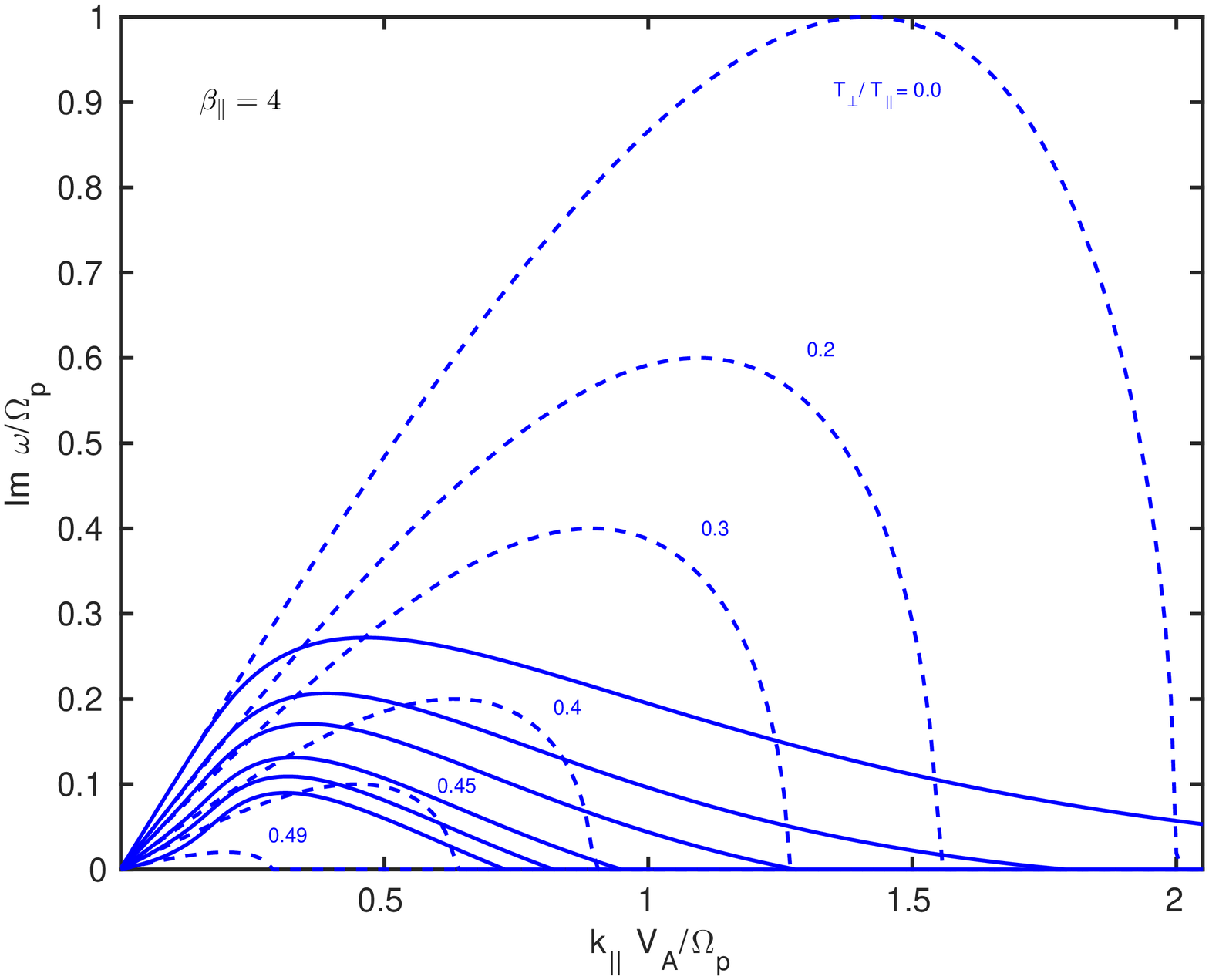}\hspace{0.03\textwidth}\includegraphics[width=0.48\linewidth]{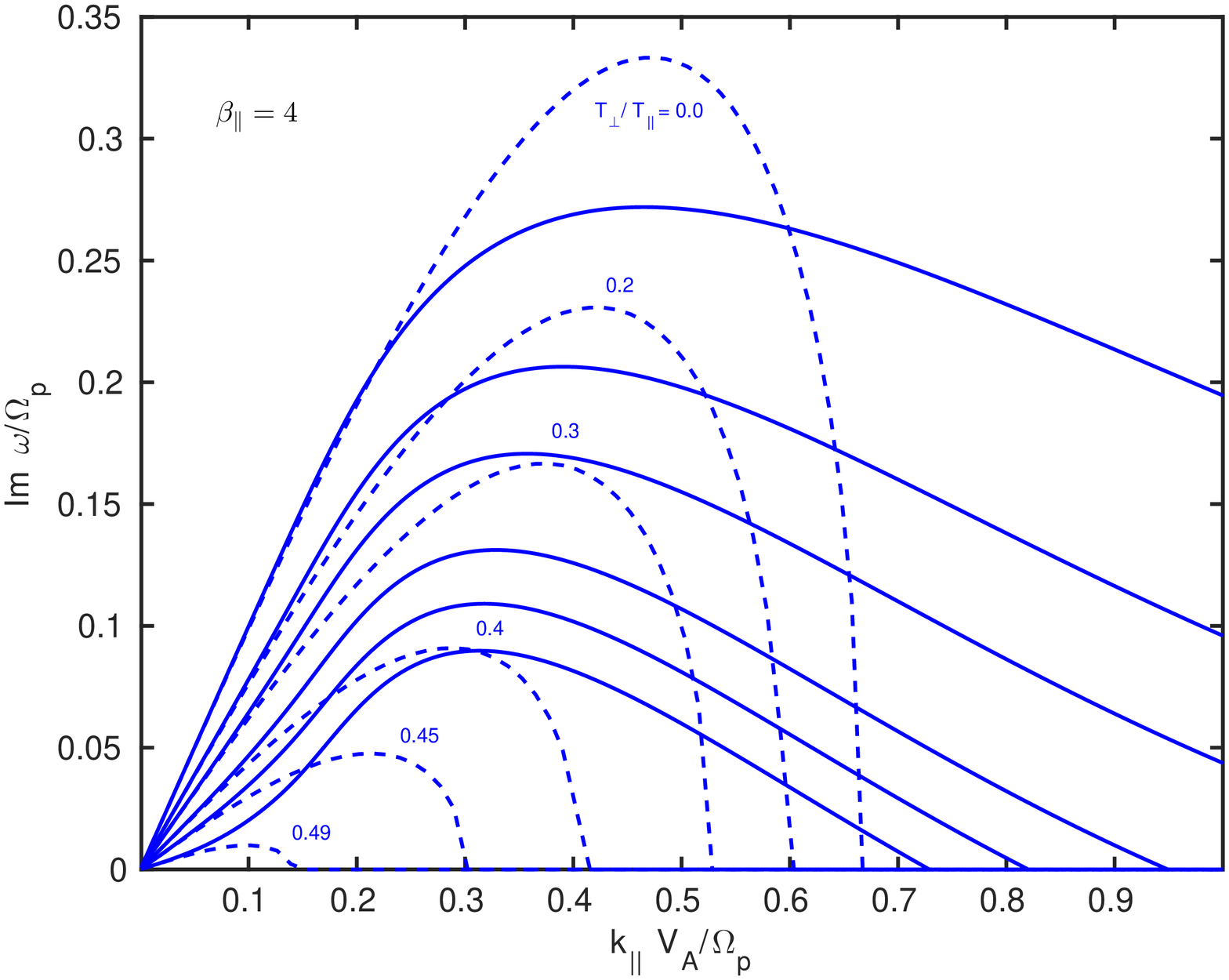}$$
  \caption{The growth rate of the parallel firehose instability of the whistler mode obtained from kinetic theory and the Hall-CGL model (left) and the refinement with FLR corrections (right).
    Similar to the previous figures but instead the frequency is plotted rather than the phase speed, and a linear scale is used to make the maximum growth rate clearly visible.
    Naturally, the growth rate (and its maximum) is not reproduced accurately by the Hall-CGL models, nevertheless, the fluid models do exhibit maxima.} 
\end{figure*}
It is useful to express the above expressions using the temperature anisotropy ratio $a_p=T_\perp/T_\parallel$ on the left hand side.
The Hall-CGL model yields 
\begin{equation} \label{eq:gmax1}
\frac{T_\perp}{T_\parallel} = 1-\frac{2}{\bpar}(1+\gamma_{max}),
\end{equation}
and the Hall-CGL-FLR1 model yields
\begin{equation} \label{eq:gmax2}
\frac{T_\perp}{T_\parallel} = 1-\frac{2}{\bpar} - \frac{\gamma_{max}}{1-\gamma_{max}}.
\end{equation}
Obviously, for $\gamma_{max}=0$, both analytic expressions (\ref{eq:gmax1}), (\ref{eq:gmax2}) yield the ``hard'' firehose threshold.
However, now as in kinetic theory, we can specify $\gamma_{max}=10^{-1}-10^{-3}$ and plot the different firehose contours, as shown in Figure 7.
\begin{figure*}
$$\includegraphics[width=0.48\linewidth]{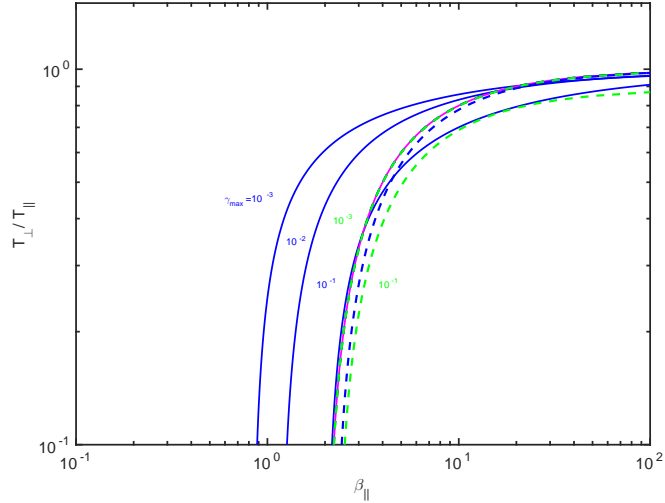}$$
  \caption{Marginally stable states for the parallel firehose instability with $\gamma_{max}=10^{-1}-10^{-3}$. The kinetic contours (blue solid lines) are from
    \cite{Hellinger2006}, Figure 1 left (Hellinger, private communication). The blue dashed lines are the marginally stable states of the Hall-CGL model and the
    green dashed lines of the Hall-CGL-FLR1 model (with first order FLR corrections). The solid magenta line is the ``hard'' firehose threshold.
    For both fluid models, the curve with $\gamma_{max}=10^{-2}$ is barely distinguishable from that with $10^{-3}$, and only the latter is plotted. Additionally,
    for both fluid models the curve with $\gamma_{max}=10^{-3}$ is indistinguishable from the hard firehose threshold. The kinetic contours
    do not match the fluid contours, except perhaps for $\bpar>10$, where both fluid models roughly match the solutions
    for $\gamma_{max}=10^{-3},10^{-2}$ and only the Hall-CGL-FLR1 model roughly matches the solution for $\gamma_{max}=10^{-1}$. Nevertheless the overall location of the fluid
    contours in the $(\bpar,T_\perp/T_\parallel)$ plane is correct.} 
\end{figure*}
The results nicely demonstrate
that a maximum growth rate is captured by the simple fluid models considered here, and that contours with a prescribed finite $\gamma_{max}$ can be considered.
The expressions are of course not precise compared to kinetic calculations. The errors in the maximum
growth rate are surprisingly large when close to the firehose threshold, i.e. when $\gamma_{max}$ is small. For example from Table 1 in \cite{Hellinger2006},
a good approximation for the parallel firehose contour with $\gamma_{max}=10^{-3}$ should be of order $T_\perp/T_\parallel = 1-0.47/(\bpar-0.59)^{0.53}$,
whereas the correction factors in (\ref{eq:gmax1}) or (\ref{eq:gmax2}) are essentially negligible for $\gamma_{max}=10^{-3}-10^{-2}$, and the parallel firehose marginally stable states
obtained using Hall-CGL models are indistinguishable and virtually equivalent to the  ``hard'' firehose threshold. For $\gamma_{max}=10^{-1}$
the fluid solutions are noticeably different from the firehose threshold but still not consistent with the kinetic contours, except perhaps for $\bpar>10$.
The kinetic contours we use here for $\gamma_{max}=10^{-2}-10^{-1}$ are the same kinetic countrours as in the left Figure 1 of \cite{Hellinger2006}
and the contour values were kindly provided to us by Petr Hellinger (private communication). 

The differences between the fluid and kinetic models can be understood from Figure 4 right, where the kinetic solution close to the threshold with $T_\perp/T_\parallel=0.49$ shows a large
``bump'' around $k V_A/\Omega_p \sim 0.1-0.5$ that deviates significantly from the fluid solution.
We do not present the maximum growth rate associated with the second order FLR corrections since the accuracy is not improved by much.
The following discussion applies only to the strictly parallel firehose instability.
The important point that is apparent from
Figure 4 right is that in the Hall-CGL model the imaginary phase speed reaches its maximum at long spatial scales or wavenumbers $k\rightarrow 0$, and for higher wavenumbers
the imaginary phase speed decreases. This means that the long wavelength limit can be used to decide if the firehose instability exists or not.
If the firehose instability does not exist at long wavelengths (or short wavenumbers) in the Hall-CGL model, it will not exist at short wavelengths. 
This implies that marginally stable states with $\gamma_{max}\rightarrow 0$ converge to the ``hard'' firehose threshold (\ref{eq:firehose}).  
The situation is different in kinetic theory where the maximum phase speed is obtained at some finite wavenumber, and the effect (i.e. the ``bump'')
is very pronounced when close to the firehose threshold. To further explore this effect, in Figure 4 right we plotted another solution with $a_p=0.501$,
i.e a solution that technically lies in a firehose-stable regime. The solution is indeed stable (with zero growth rate) at small wavenumbers both in the kinetic
theory and in the fluid model. However, the difference is that while the fluid model remains stable for higher wavenumbers, the kinetic solution still develops
the firehose instability. The consequence of this is that kinetic marginally stable states with small $\gamma_{max}$ can lie below the ``hard'' CGL firehose
threshold, further implying that the kinetic firehose instability can exist at finite wavenumbers for plasma beta values slightly below $\bpar=2$. That kinetic
contours for $\gamma_{max}=10^{-3},10^{-2}$ can fall below $\bpar=2$ is clearly visible in Figure 7
(also in Figure 1 of \cite{Hellinger2006} or Figure 2 of \cite{Bale2009}). 

We note that the parallel firehose instability for quasi-parallel propagation angles (not strictly parallel) in the Hall-CGL model (and models with FLR corrections)
can also reach its maximum
imaginary phase speed at a finite range of wavenumbers (i.e. the imaginary phase speed can be higher than the phase speed obtained in the long-wavelength limit), and
this is briefly addressed in the next section. The parallel firehose instability usually reaches its maximum growth rate at $\theta=0^\circ$.

To clearly show that for a high plasma $\bpar$ the stabilization mechanism is mainly due to the FLR corrections and not the Hall term, in Figure 8 we plotted
the parallel firehose instability for $\bpar=100$. It is shown that even though the Hall term in the Hall-CGL model (blue dashed lines) eventually stabilizes the
firehose instability at sufficiently high wavenumbers, the solutions are very far from the kinetic solutions. In contrast, the Hall-CGL-FLR1 model (green dashed lines)
stabilizes the solutions reasonably well. Interestingly, for such a high plasma beta, the strong ``bump'' that was shown in the kinetic
imaginary phase speed for $\bpar=4$ (Figure 4 right, when close to the firehose threshold) is not present here.
This implies that it is now possible to compare solutions that are close to the
firehose threshold, which is done in the Figure 8 right. It shows that the Hall-CGL-FLR1 model captures the parallel kinetic growth rate extremely well.
\begin{figure*}
$$\includegraphics[width=0.455\linewidth]{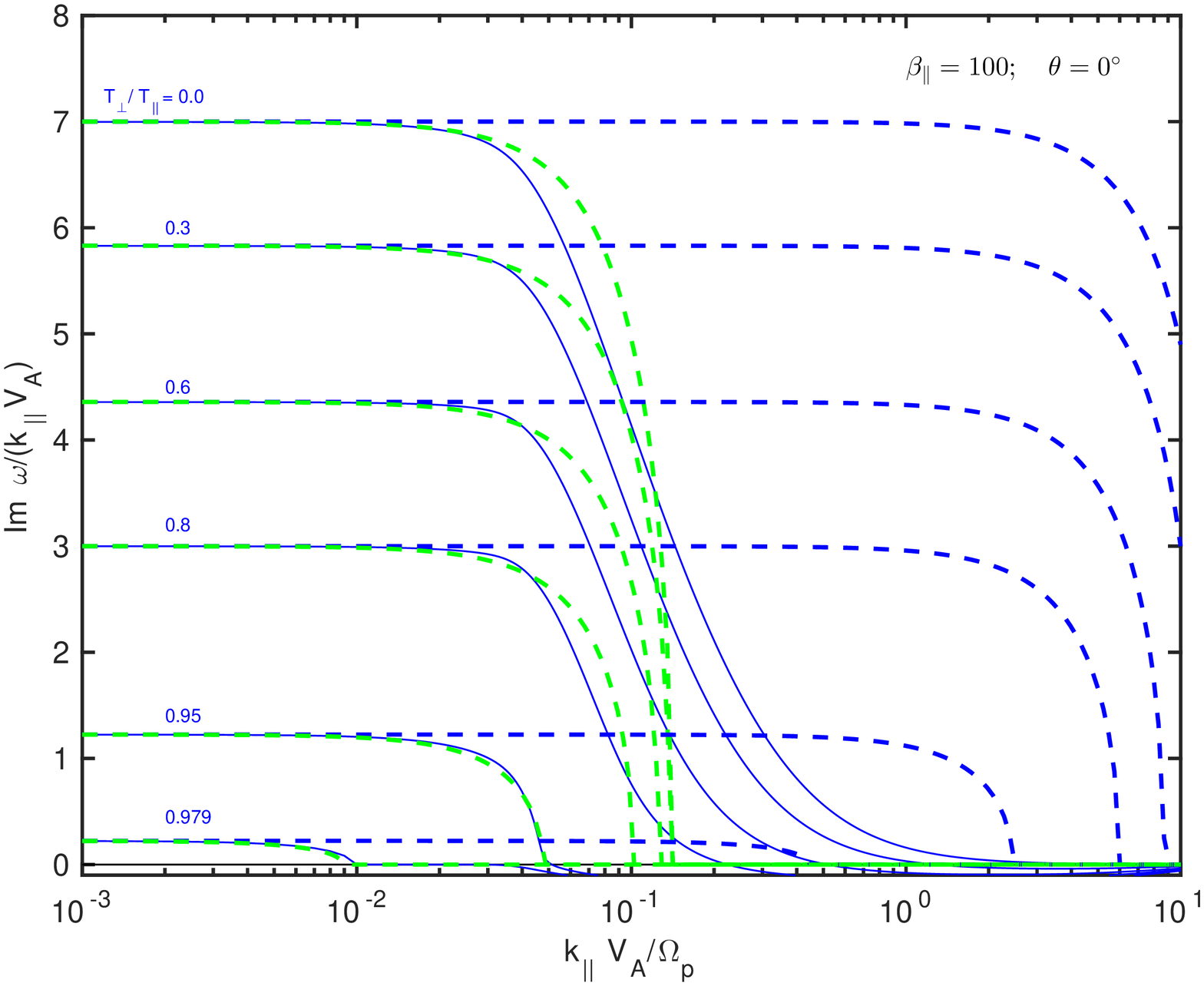}\hspace{0.03\textwidth}\includegraphics[width=0.48\linewidth]{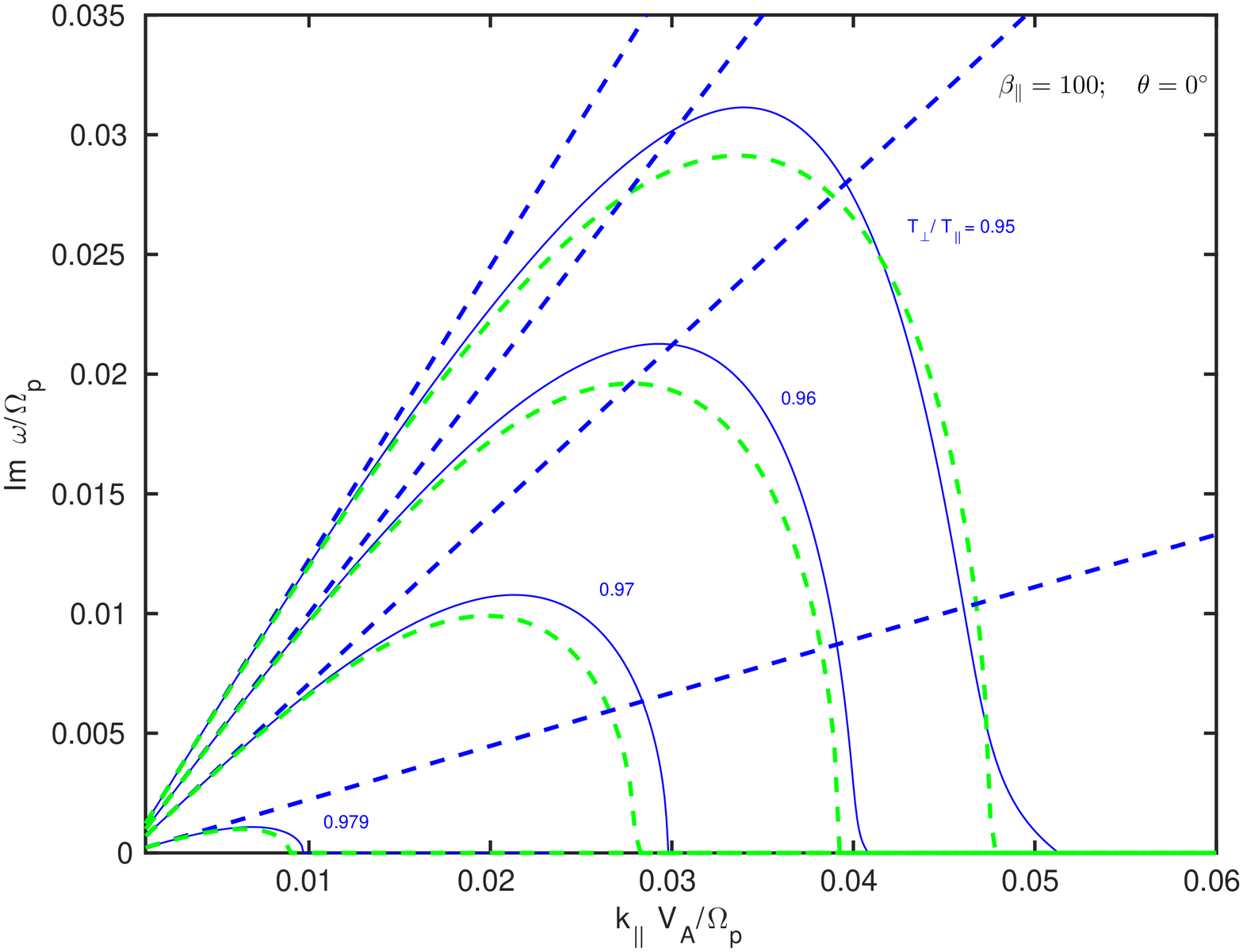}$$
  \caption{The parallel firehose instability for $\bpar=100$. Kinetic solutions are solid lines, the dashed blue lines are solutions of the Hall-CGL model,
  and the dashed green lines of the Hall-CGL-FLR1 model. It shows that (for strictly parallel propagation), the stabilization mechanism is due mainly to the FLR corrections.} 
\end{figure*}

We investigated kinetic solutions for different plasma $\bpar$ values, concentrating on solutions close to the firehose threshold, with the aim of
exploring the occurrence of the ``bump'' in the imaginary phase speed. Our results indicate (not shown), that the ``bump'' in the kinetic imaginary phase speed,
i.e. the strong enhancement of the growth rate, exists for all plasma beta values up to roughly $\bpar=20$. We therefore conclude that the significant
differences for the parallel firehose instability growth rate when close to the firehose threshold, exists for all $\bpar<20$.
Only for $\bpar>20$ do the growth rates obtained from the kinetic description and from the simple fluid models considered here, start to converge for small $\gamma_{max}$.
This conclusion is consistent with the finding of Figure 7. 

\section{The Oblique firehose instability}
The dispersion relation for the Hall-CGL model with FLR corrections can be written in the convenient form
\begin{eqnarray}
  &&  \Big(\widetilde{\omega}^2-\widetilde{k}_\parallel^2 \widetilde{v}_{A\parallel}^2\Big)\Big(\widetilde{\omega}^4-A_2\widetilde{\omega}^2+A_0\Big) =
  \widetilde{k}^2 \widetilde{k}_\parallel^2 \Big[ \widetilde{\omega}^4-\bpar(\frac{3}{2}\widetilde{k}_\parallel^2+a_p \widetilde{k}_\perp^2)\widetilde{\omega}^2
    + \widetilde{k}_\parallel^2 \widetilde{k}_\perp^2 \frac{5}{4}a_p\bpar^2\Big] + \mathcal{P}^{\textrm{FLR}}; \label{eq:HallCGLdisp}\\
 &&   A_2 = \widetilde{k}_\perp^2 \left(1+a_p\bpar\right) + \widetilde{k}_\parallel^2 \left(\widetilde{v}_{A\parallel}^2+\frac{3}{2}\bpar\right); \nonumber \\
 && A_0 = \frac{3}{2}\widetilde{k}_\parallel^2 \bpar \left[ \widetilde{k}_\parallel^2 \widetilde{v}_{A\parallel}^2  + 
\widetilde{k}_\perp^2 \left(1+a_p\bpar-\frac{1}{6}a_p^2\bpar\right)  \right]. \label{eq:CGLparameters}
\end{eqnarray}
The left hand side represents the dispersion of the CGL model and the right hand side the Hall and FLR contributions that couple the CGL modes and make
the solutions length-scale dependent. The Hall-CGL model is obtained by setting
$\mathcal{P}^{\textrm{FLR}}=0$ and the polynomial representing the FLR1 contributions is written down in the Appendix. The dispersion relation is written in
normalized wavenumbers and frequencies as discussed previously and the parallel Alfv\'en speed is normalized as
$\widetilde{v}_{A\parallel}^2={v}_{A\parallel}^2/V_A^2=1+\frac{\bpar}{2}(a_p-1)$. For isotropic temperatures $a_p=1$, the CGL model is not equivalent
to the MHD model and naturally, the Hall-CGL model is also not equivalent to the Hall-MHD model. 
Here we first explore the growth rate of the oblique firehose instability for
propagation angles of $\theta=30^\circ$ and $\theta=60^\circ$ with respect to the mean magnetic field, and for the present we ignore the parallel firehose
instability that also exists for the first angle of propagation. The solutions are shown in Figure 9 (for the Hall-CGL model) and Figure 10
(for the Hall-CGL-FLR1 model). 
We find that again, the CGL model accurately describes the growth rate (the imaginary phase speed is shown) of the oblique firehose
instability at sufficiently long spatial scales and that the stabilization is due to the Hall term and the FLR corrections. However, as before, we find
that the kinetic solution can become firehose unstable at higher wavenumbers even if the kinetic solution is stable for small wavenumbers,
explaining the difference between the CGL ``hard'' threshold and the kinetic solutions obtained for a small $\gamma_{max}$.
We do not attempt to derive here analytically the Hall-CGL oblique firehose marginally stable states for a prescribed $\gamma_{max}$, since all the modes are strongly coupled
through the dispersion relation (\ref{eq:HallCGLdisp}) and we do not see any obvious way to separate them. Only in the somewhat ``academic'' limit of
$a_p=0$ (i.e. $T_\perp=0$ and $\bpar$ arbitrary) one mode that connects to the parallel ion-acoustic mode can be separated, having the dispersion relation
$\widetilde{\omega}=\widetilde{k}\cos\theta\sqrt{\frac{3}{2}\bpar}$,
and the Alfv\'en/ion-cyclotron mode and the whistler mode remain coupled, but this limit is not pursued further here. 
Additionally, Figures 9 and 10 reveal clearly 
that for a small $\gamma_{max}$, the corrections to the hard firehose threshold will be basically negligible (as was shown for the parallel propagation) and
the marginally stable states will not match those obtained from kinetic theory, especially for a very small $\gamma_{max}$. 
\begin{figure*}
$$\includegraphics[width=0.48\linewidth]{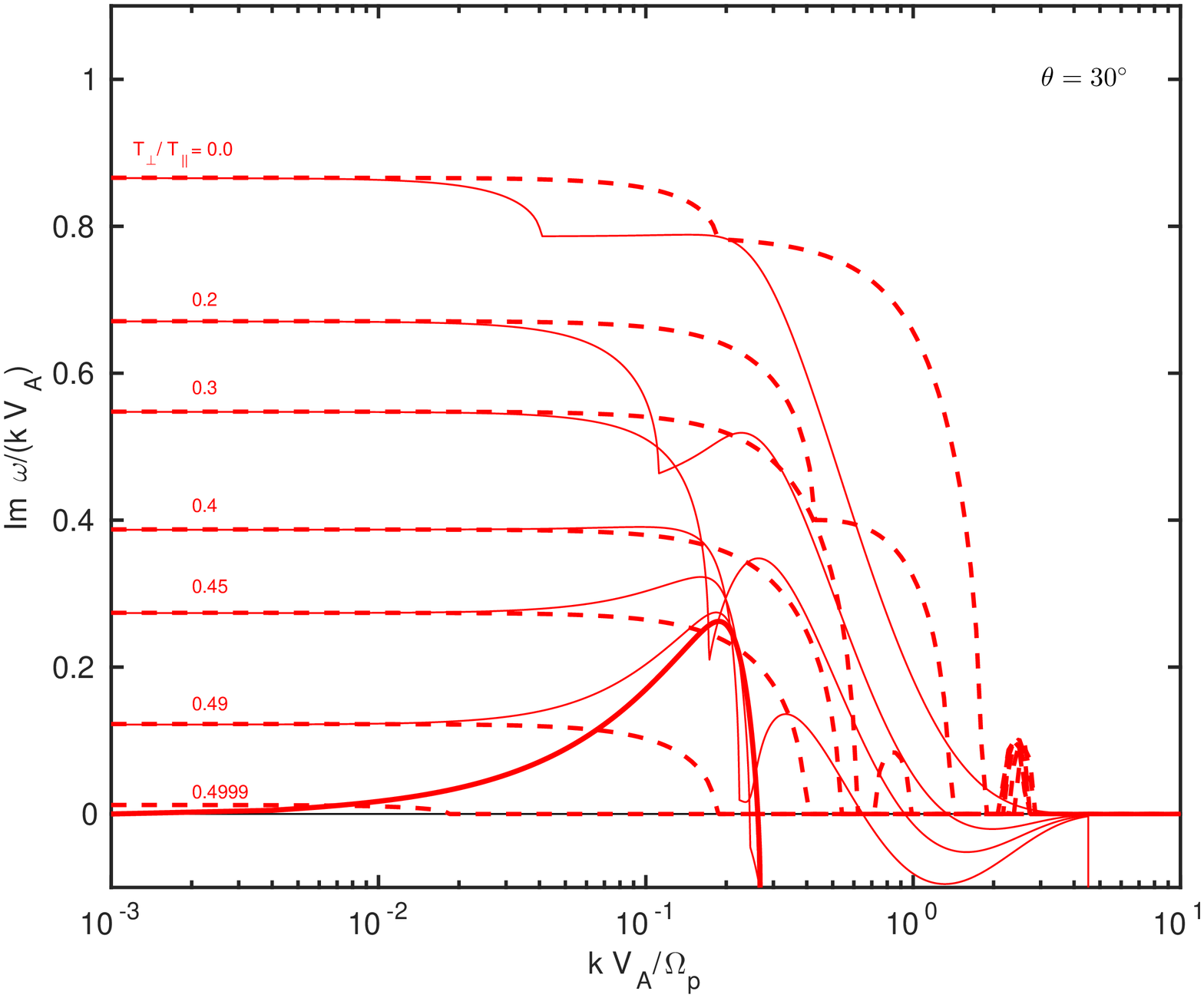}\hspace{0.03\textwidth}\includegraphics[width=0.48\linewidth]{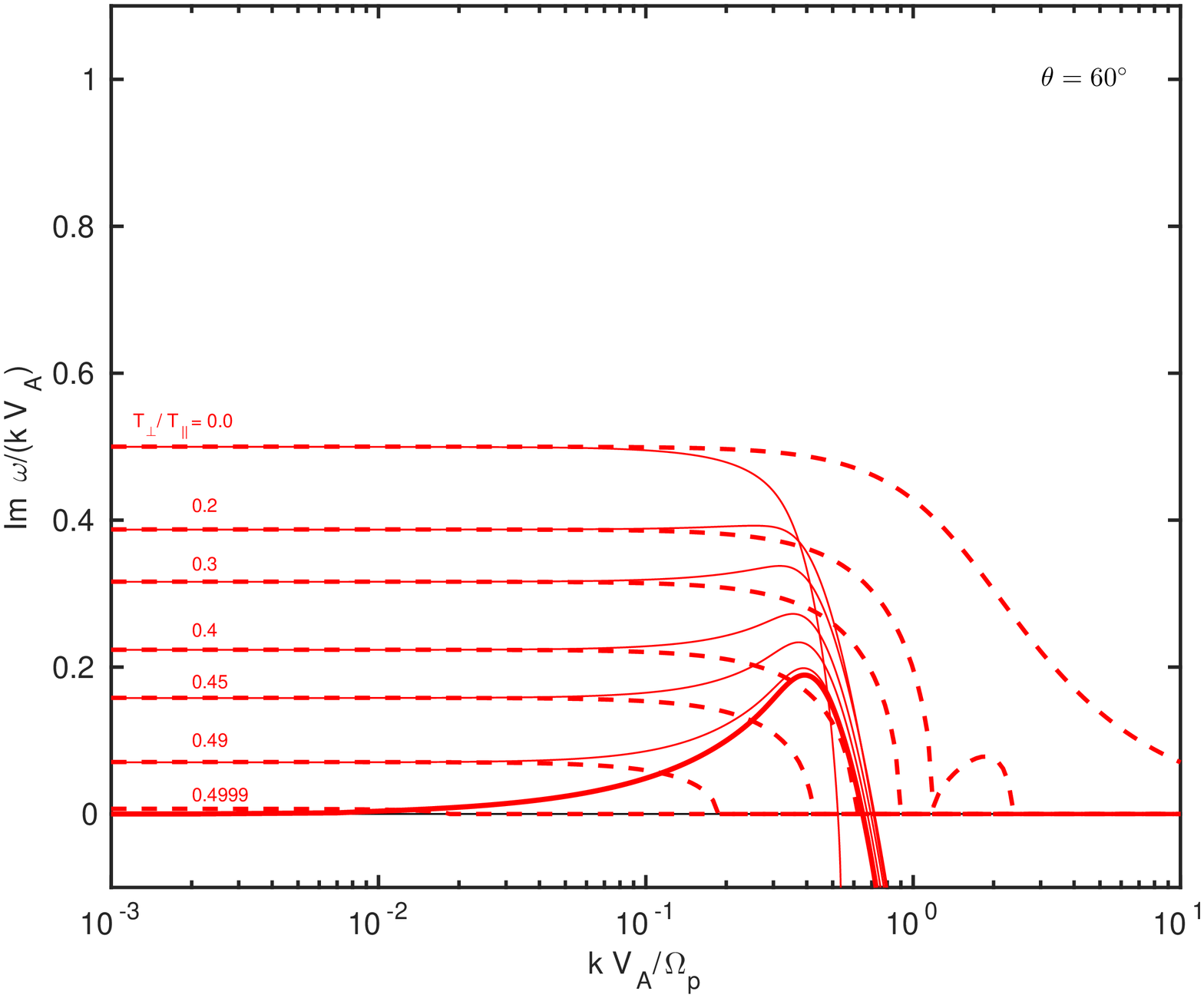}$$
  \caption{The oblique firehose instability of the Alfv\'en mode. Left: propagation angle $\theta=30^{\circ}$; Right: propagation angle $\theta=60^\circ$.
    Same as previous figures, $\bpar=4$ and the temperature anisotropy is varied as $T_\perp/T_\parallel=0.49; 0.45; 0.4; 0.3; 0.2; 0.0$.
    Also two solutions with $T_\perp/T_\parallel=0.4999$ that are very close to the threshold are plotted and are emphasized with a thicker curve. 
    Solid lines are solutions of kinetic theory and dashed lines are solutions of the Hall-CGL model. The left figure shows that there is an erroneous
    second instability in the Hall-CGL model around $k V_A/\Omega_p \sim 2-3$ (a small bump) for solutions with $T_\perp/T_\parallel \le 0.3$,
    including the solution very close to the threshold and also another instability slightly below $k V_A/\Omega_p=1$ (for parameter $T_\perp/T_\parallel=0.3$).
    In the right figure the solution with $T_\perp/T_\parallel=0.0$ is not fully stabilized and the solution with $T_\perp/T_\parallel=0.2$ has a secondary
  instability. These erroneous instabilities are removed by the simple FLR corrections, which is shown in the next figure.} 
\end{figure*}
\begin{figure*}
$$\includegraphics[width=0.48\linewidth]{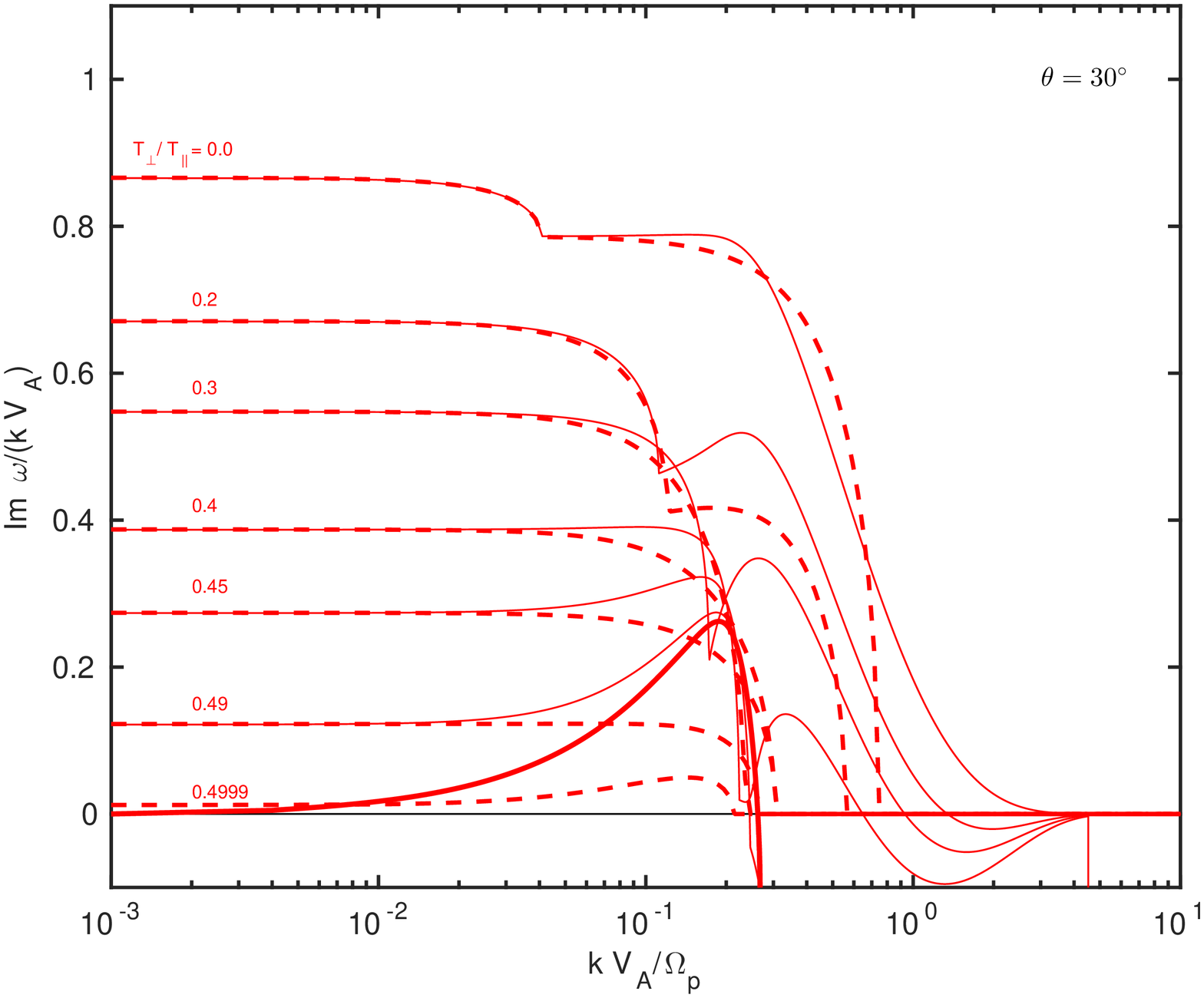}\hspace{0.03\textwidth}\includegraphics[width=0.48\linewidth]{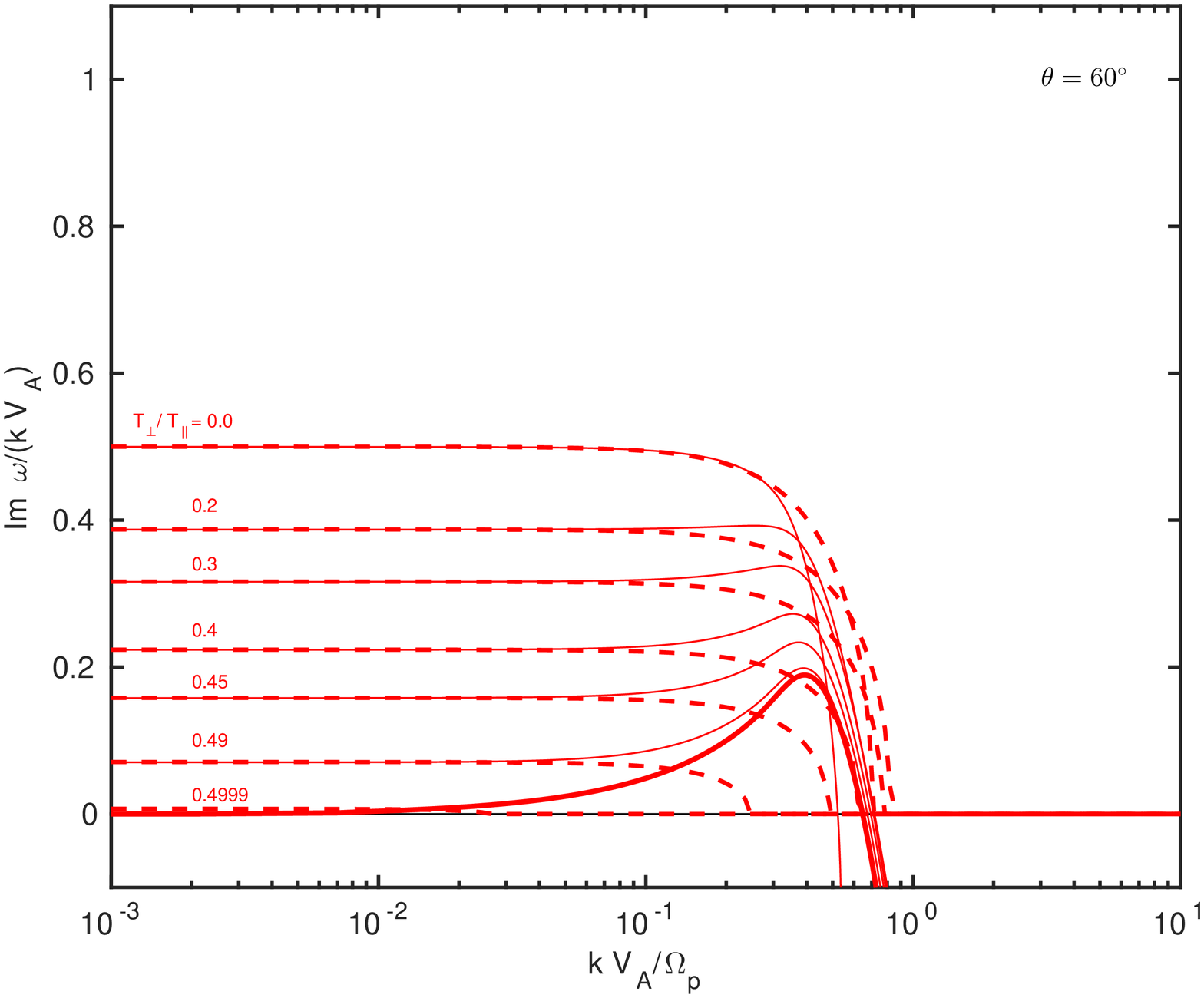}$$
  \caption{The oblique firehose instability of the Alfv\'en mode. Same as previous figure, but the CGL model is extended to include simple
    FLR1 corrections. The stabilization is generally improved, especially for solutions far from the threshold with $T_\perp/T_\parallel=0.0; 0.2$.
    Most importantly, the erroneous secondary instabilities found in the Hall-CGL model for some cases are completely removed.} 
\end{figure*}

Interestingly, Figure 9 left shows that the Hall-CGL model exhibits an instability around $k V_A/\Omega_p\sim 2-3$ that is not observed in the kinetic solutions.
The instability might appear unimportant here (a small bump) because the imaginary phase speed is plotted, but, the growth rate is quite large and the
instability is quite strong. This erroneous instability that is present in the Hall-CGL model is completely removed by considering the FLR corrections, as
shown in Figure 10. Figure 10 left (\& 9 left) shows a curious sharp transition in the growth rate for solutions with $a_p=0.0$ and $a_p=0.2$
at relatively long wavelengths (around $kV_A/\Omega_p=0.04$ and $0.12$).
These transitions are easily explained by simultaneously plotting the oblique and the parallel firehose instability, as is done in Figure 11, where the
Hall-CGL-FLR1 model is used. Figure 11 shows that the sharp transitions are caused by the ``interaction'' of the parallel and oblique firehose instability,
where at some critical wavenumber, their growth rates match. At long wavelengths (small wavenumbers) up to this critical point both instabilities have
zero real frequency and the figure therefore shows an example of a non-propagating parallel firehose instability. Figure 11 also shows that the (obliquely propagating)
parallel firehose instability can reach a maximum imaginary phase speed at some finite wavenumber in a fluid model as well, i.e. the imaginary phase speed can be higher
than the one obtained as $k\to 0$. Notably, the case $a_p=0$ shows that at long wavelengths, the parallel firehose phase speed and
the ``interaction point'' are reproduced very accurately by the fluid model. However, this is not true for the case $a_p=0.2$, where the parallel firehose
phase speed is not captured accurately. This is a consequence of the erroneous $1/6$ factor in the dispersion relation of the CGL model (in the quantity $A_0$)
that is also responsible for the well-known erroneous mirror threshold. The term containing this factor is proportional to $\frac{1}{6}a_p^2\bpar$ and,
for $a_p=0$ the term vanishes, resulting in the correct long wavelength limit. 
\begin{figure*}
$$\includegraphics[width=0.48\linewidth]{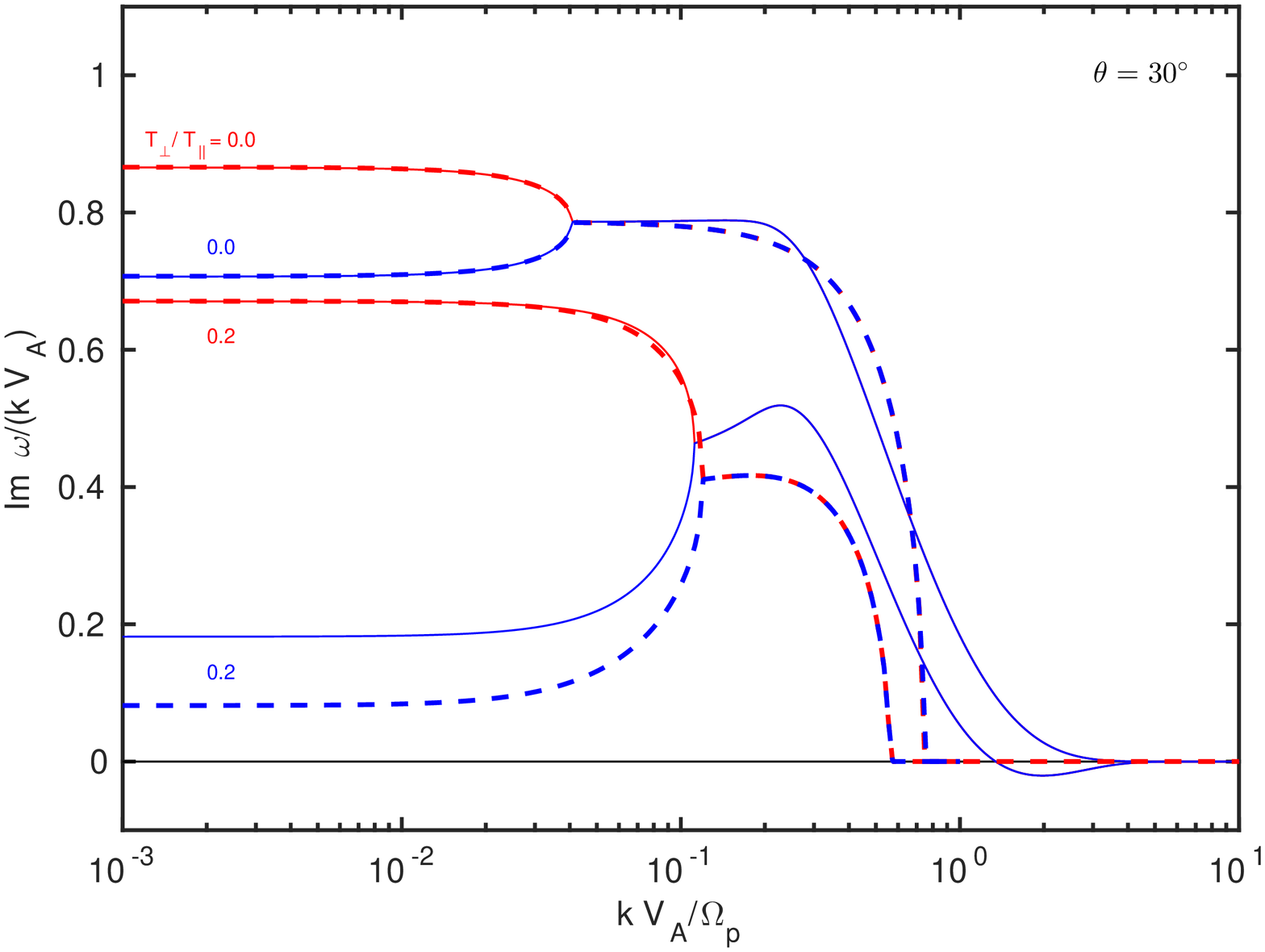}$$
  \caption{The sharp transitions in the growth rate of the oblique firehose instability (red)
    visible in the previous figure for $\theta=30^\circ$ and $T_\perp/T_\parallel=0.0; 0.2$ can be clarified by also plotting the parallel firehose instability (blue),
    as is done here. The kinetic solutions are solid lines and solutions of the Hall-CGL-FLR1 model correspond to dashed lines. The parallel and oblique firehose instability
    ``interact'' with each other and at some critical wavenumber ($\sim 0.04$ for the first case and $\sim 0.12$ for the second case) their growth rates match.
    At small wavenumbers up to this critical point both instabilities have frequencies that are purely imaginary. This is an example of a parallel
    firehose instability that is non-propagating.} 
\end{figure*}

Here we now explore the maximum growth rate for oblique propagation. We now concentrate on solutions close to the firehose threshold, i.e. when
$\gamma_{max}$ is relatively small, since these solutions determine what waves (at what angle) will be preferentially excited when the firehose threshold
is reached. Figure 12 shows kinetic solutions for the growth rate of the oblique firehose instability (red lines) with $\bpar=4, a_p=0.49$ for various angles of
propagation between $\theta=10^\circ-80^\circ$. The maximum growth rate of the oblique firehose instability is reached around $\theta=55^\circ$.
Two solutions for the parallel firehose instability (blue lines) for $\theta=0^\circ$ and $\theta=5^\circ$ are also plotted. 
The parallel firehose instability reaches a slightly higher maximum growth rate than the oblique firehose instability. Figure 13 left shows Hall-CGL-FLR1 solutions
for the same parameters. The plots show that even though the solutions are physically plausible and reasonably well-behaved, there are significant
differences in comparison to the kinetic calculation. For example, the oblique firehose instability reaches its maximum growth rate around $\theta=35^\circ$ and
the maximum growth rate of the parallel firehose instability is lower than that of the oblique firehose instability.
Importantly, the fluid solutions are stabilized at smaller wavenumbers and also reach significantly smaller values of $\gamma_{max}$, i.e. the fluid solutions 
lie closer to the firehose threshold than the kinetic solutions. This is not surprising and is fully expected from our previous discussion
of solutions close to the firehose threshold (e.g. Figures 10, 7, 6 right, 4 right). It can therefore be easily argued that the differences originate mainly
from the fluid model being closer to its threshold than the kinetic model, but that the physical stabilization mechanisms are in fact very similar.
Without much fine-tuning, we have therefore chosen to move the fluid model a little further from the threshold by prescribing $a_p=0.4$ and the results are plotted
in Figure 13 right. Figure 13 right is very similar to the kinetic solutions shown in Figure 12.
The oblique firehose instability reaches its maximum around $\theta=50^\circ$, and the parallel firehose instability growth rate is slightly higher than
that of the oblique firehose. The values of $\gamma_{max}$ and the wavenumbers where $\gamma_{max}$ occurs are close to the kinetic calculation
for both instabilities. This is a remarkable result for such a simple fluid model.  
\begin{figure*}
$$\includegraphics[width=0.48\linewidth]{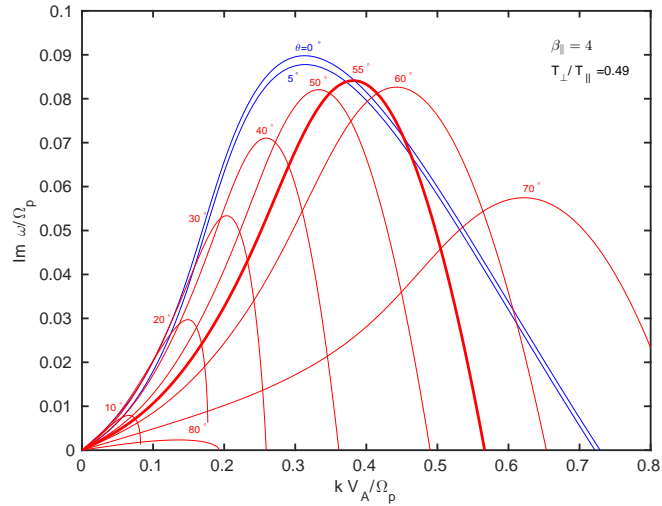}$$
  \caption{Kinetic solutions. The growth rate for various angles of propagation for the oblique firehose instability of the Alfv\'en mode (red lines).
    The parameters are $\bpar=4$ and $T_\perp/T_\parallel=0.49$. The maximum growth rate is reached for $\theta=55^{\circ}$, and the solution is emphasized with
    a thicker line. For comparison, two solutions for the parallel firehose instability of the whistler mode (blue lines) are plotted.} 
\end{figure*}
\begin{figure*}
$$\includegraphics[width=0.48\linewidth]{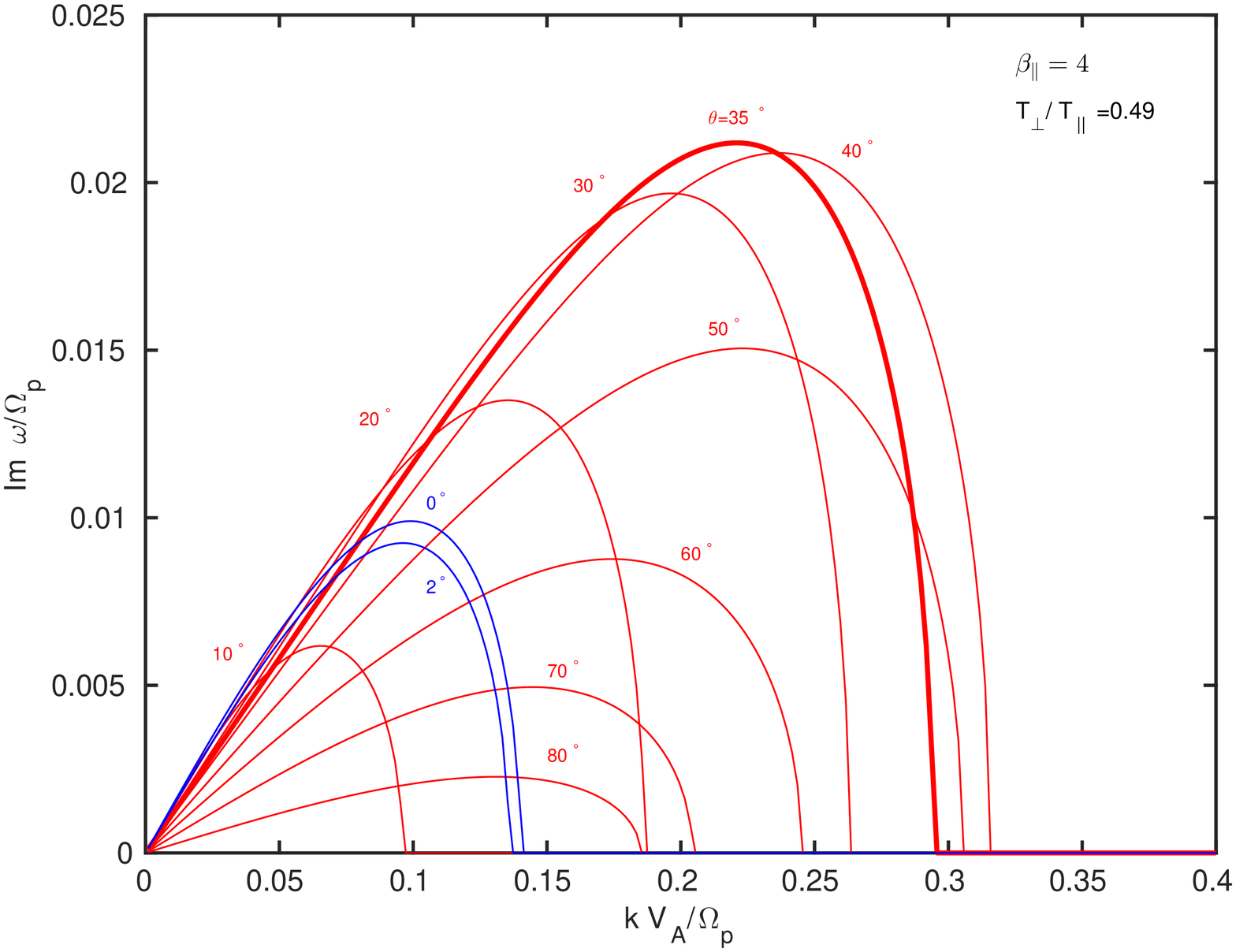}\hspace{0.03\textwidth}\includegraphics[width=0.48\linewidth]{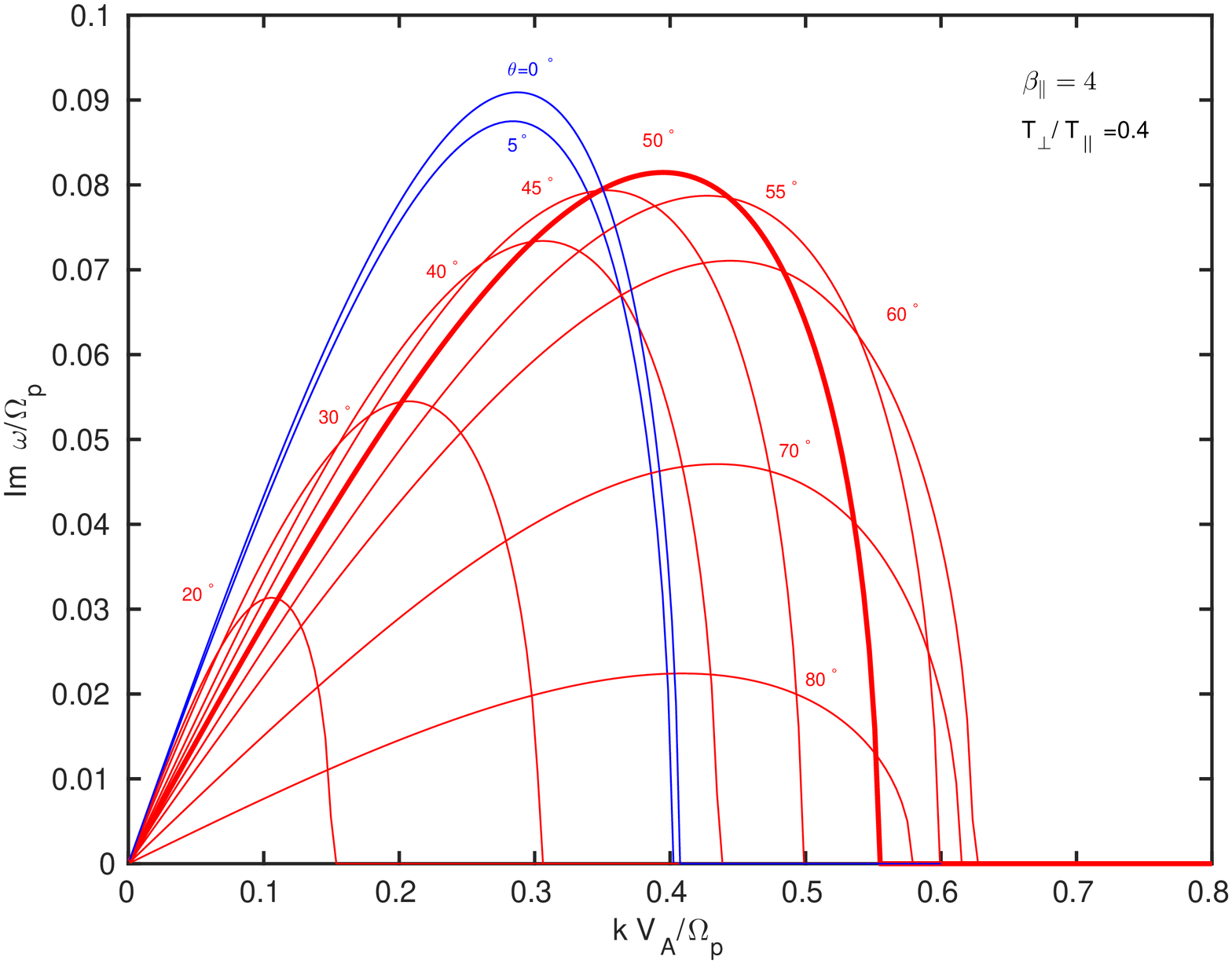}$$
  \caption{Hall-CGL-FLR1 solutions. The growth rate of the oblique firehose instability (red lines) and the parallel firehose instability (blue lines),
    for various propagation angles. Left: $T_\perp/T_\parallel=0.49$, the oblique firehose reaches a maximum growth rate at $\theta=35^{\circ}$,
    and the parallel firehose has a smaller maximum growth rate. However, it can be argued that the discrepancy arises because the kinetic solutions
    are further away from their kinetic threshold. Right: $T_\perp/T_\parallel=0.40$, meaning the fluid model is moved a little further beyond the threshold.
    The oblique firehose reaches maximum growth rate at $\theta=50^\circ$ and the parallel firehose growth rate is higher. The solutions are quite close
    to the kinetic solutions from Figure 11, which is true for the angle where the oblique firehose reaches its maximum, the actual values of the
    maximum growth rate and also that the parallel firehose has still slightly higher growth rate.} 
\end{figure*}

The previous results can be better visualized by plotting the maximum growth rate in the $(k,\theta)$ plane and creating contour plots.
Figure 14 uses the Hall-CGL-FLR1 model and shows contour plots for the same parameters used in Figure 13.
The solutions have already been discussed in detail. Nevertheless, the figure is a nice visualization of the parallel and oblique firehose instability and
demonstrates, that close to the firehose threshold, the parallel firehose instability is strongly suppressed.
Solutions for the Hall-CGL model using identical parameters are shown in Figure 15. The parallel firehose instability is
enhanced for both cases, and is not suppressed when close to the threshold. This illustrates that the FLR corrections are responsible for 
the suppression of the parallel firehose instability when close to the firehose threshold. 
\begin{figure*}
$$\includegraphics[width=0.48\linewidth]{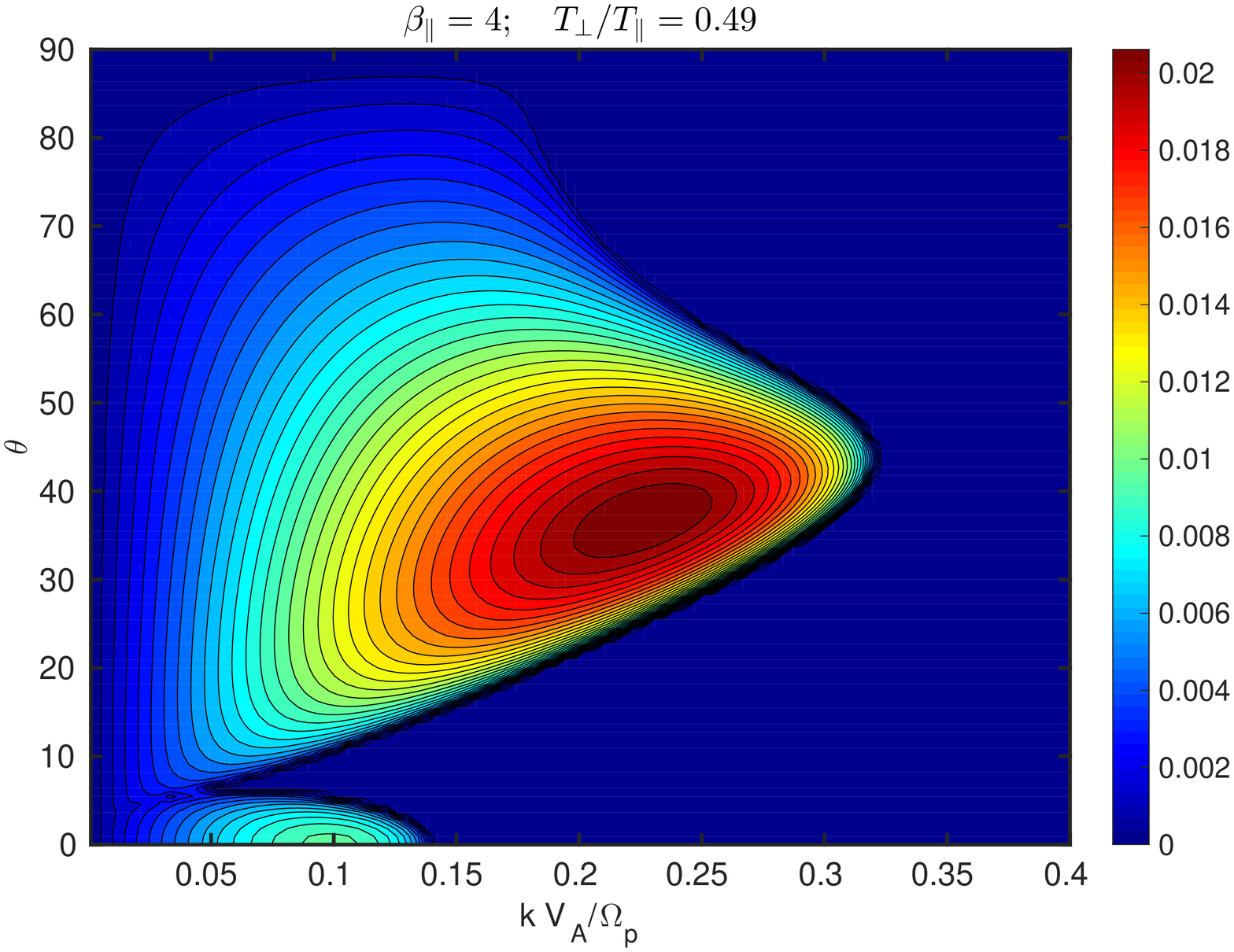}\hspace{0.03\textwidth}\includegraphics[width=0.48\linewidth]{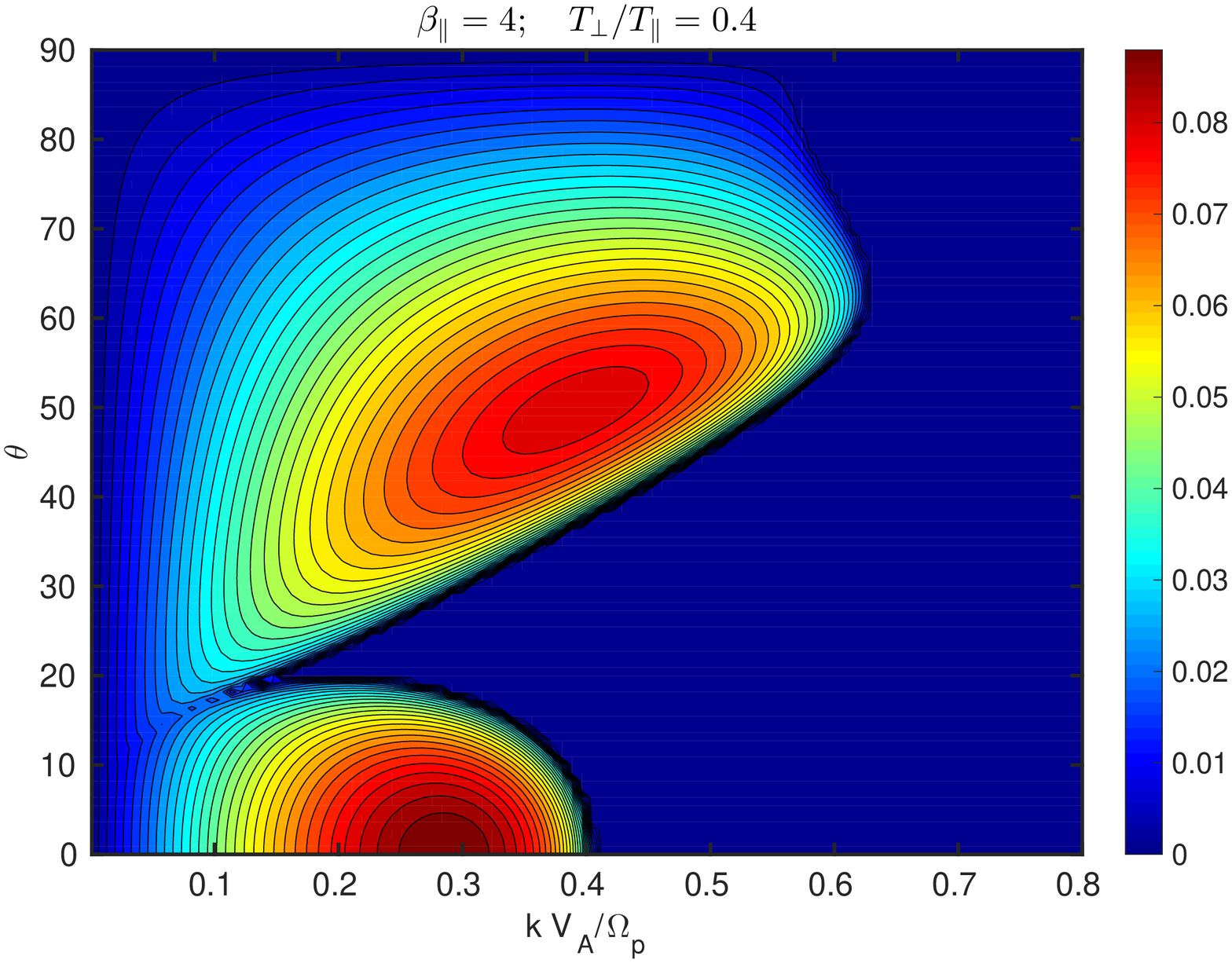}$$
  \caption{The maximum growth rate  plotted in the $(k,\theta)$ plane. The parameters are the same as in Figure 13, i.e.,
    the fluid model is Hall-CGL-FLR1, and the parameters are $\bpar=4$, $T_\perp/T_\parallel=0.49$ (left), $T_\perp/T_\parallel=0.40$ (right).
    The Figure nicely shows the parallel and oblique firehose instability. It also shows that close to the firehose threshold (left figure),
    the parallel firehose instability is strongly suppressed in the Hall-CGL-FLR1 fluid model.} 
\end{figure*}
\begin{figure*}
$$\includegraphics[width=0.48\linewidth]{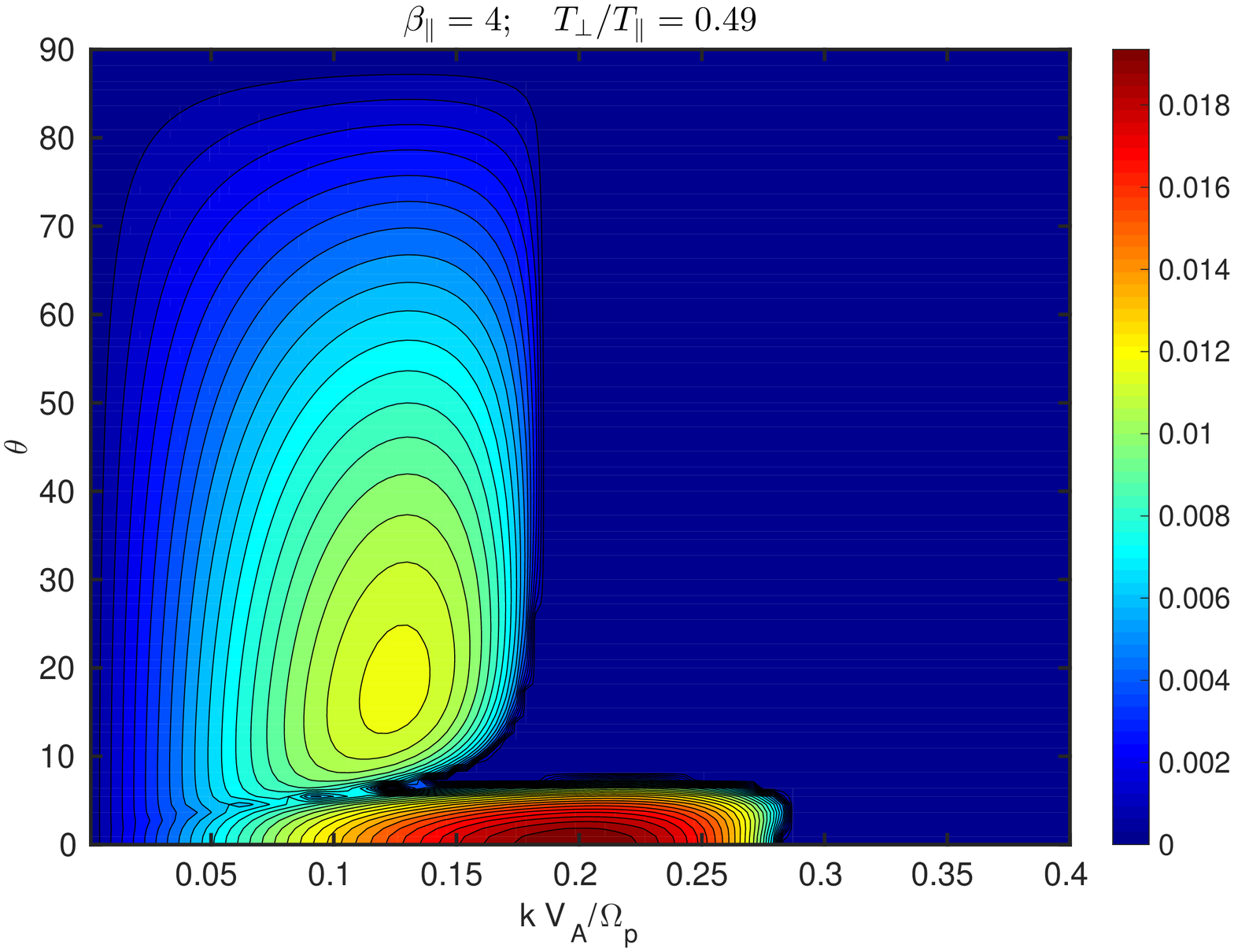}\hspace{0.03\textwidth}\includegraphics[width=0.48\linewidth]{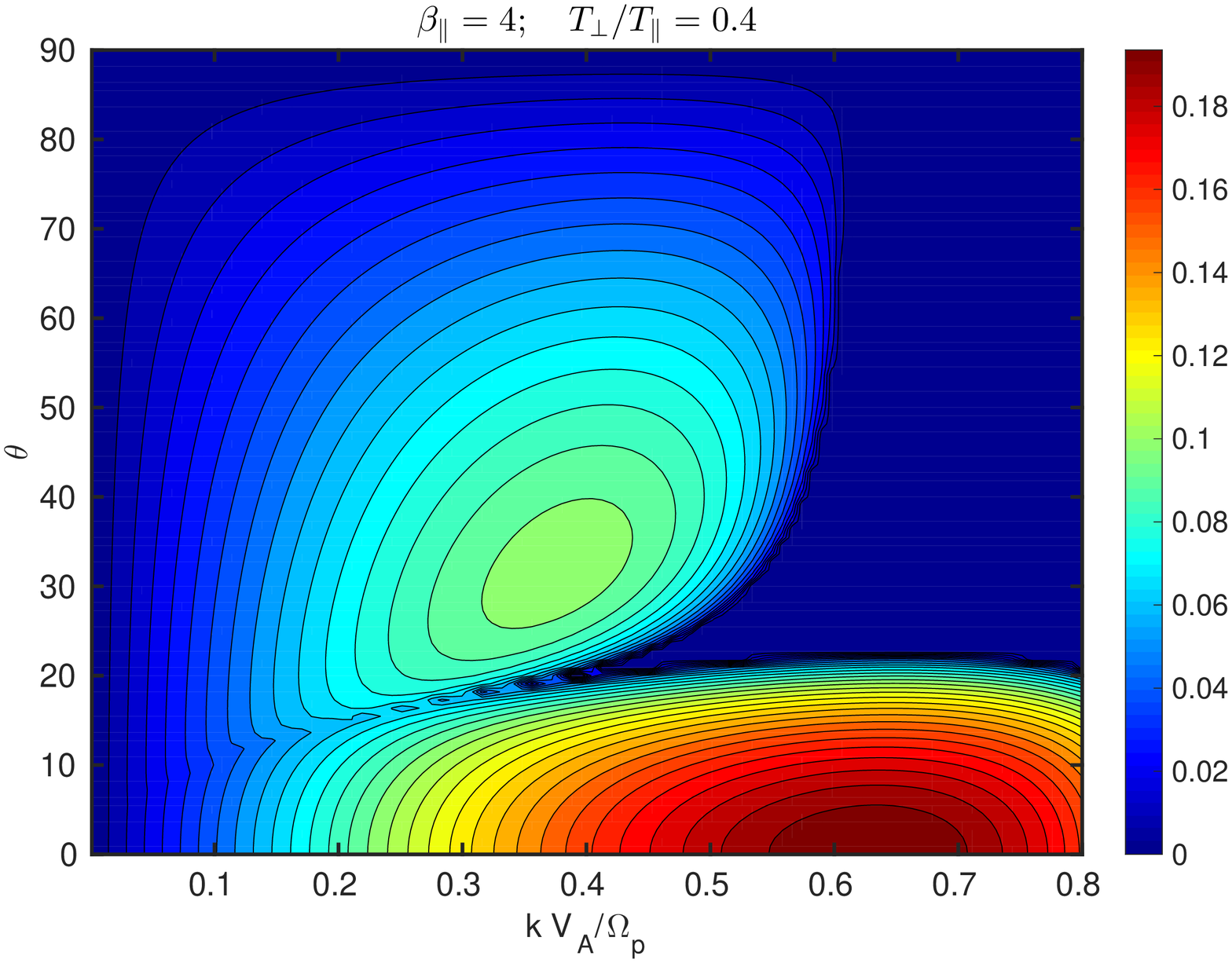}$$
  \caption{The same parameters and the same range of wavenumbers as in Figure 14, but the Hall-CGL model is plotted, i.e. compared to Figure 14, the
    FLR corrections are turned off. In both cases, the parallel firehose instability is strongly enhanced.} 
\end{figure*}

It is useful to evaluate the Hall-CGL and Hall-CGL-FLR1 models exactly at the firehose threshold, and the results are shown in Figure 16.
The Hall-CGL model is stable at small wavenumbers, but develops a spurious instability around $kV_A/\Omega_p\sim 2$, as
already discussed in Figure 9. Inclusion of FLR corrections eliminates this spurious instability and the resulting system has a 
weak oblique firehose instability with maximum growth rate $\gamma_{max}\sim 7\times 10^{-3}$ and direction of propagation around $\theta=30^\circ$.   
In the Hall-CGL and Hall-CGL-FLR1 models, the strictly parallel ($\theta=0^\circ$) firehose instability must completely vanish when evaluated directly
at the firehose threshold, which is easy to see from the analytic solutions (\ref{eq:Hall-wU}), (\ref{eq:FLR-w1}). For oblique propagation, we do not
see any obvious way to analytically show that the Hall-CGL-FLR1 model always yields an oblique firehose instability when evaluated
close to the threshold, and the system can be explored only numerically. 
\begin{figure*}
$$\includegraphics[width=0.48\linewidth]{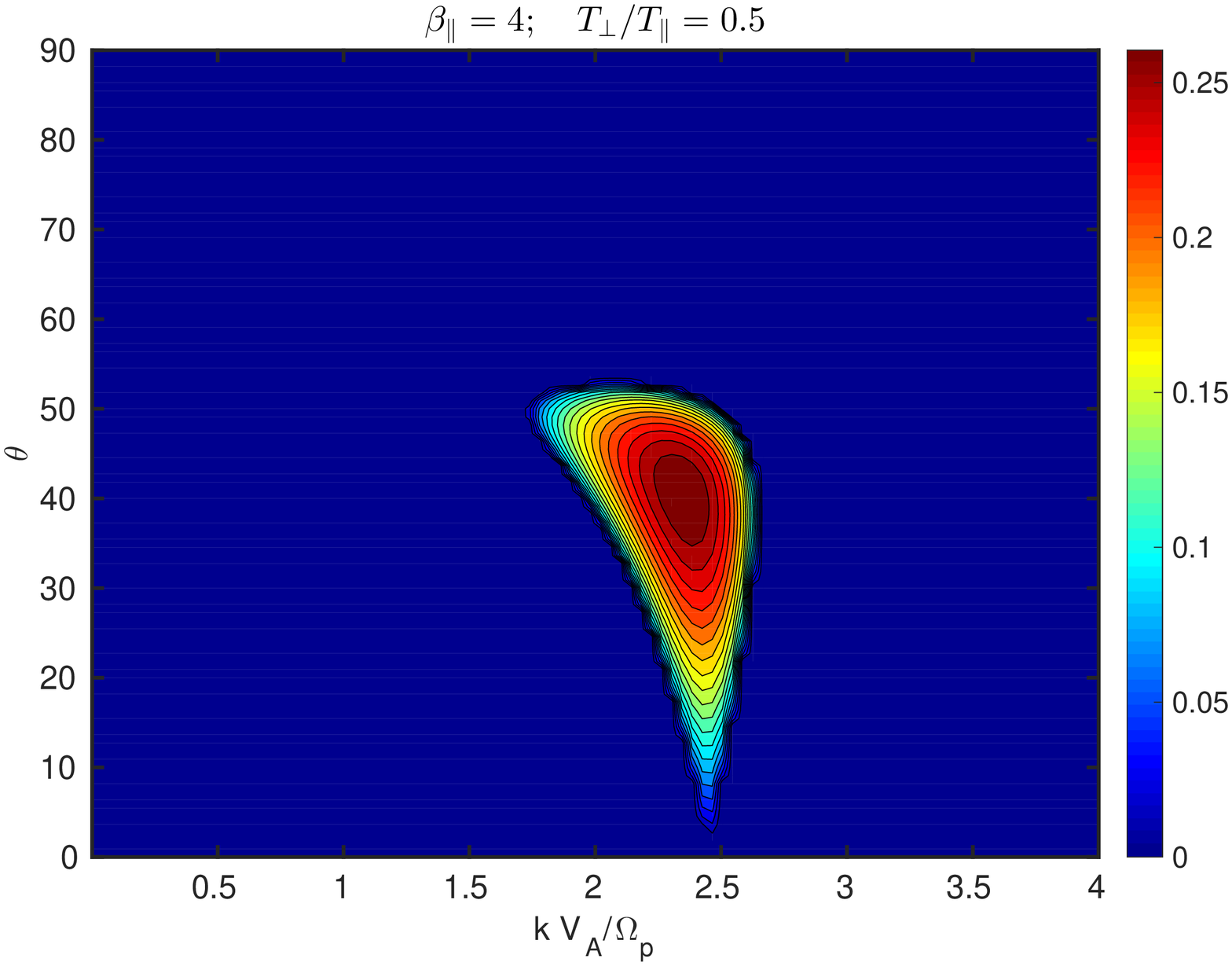}\hspace{0.03\textwidth}\includegraphics[width=0.48\linewidth]{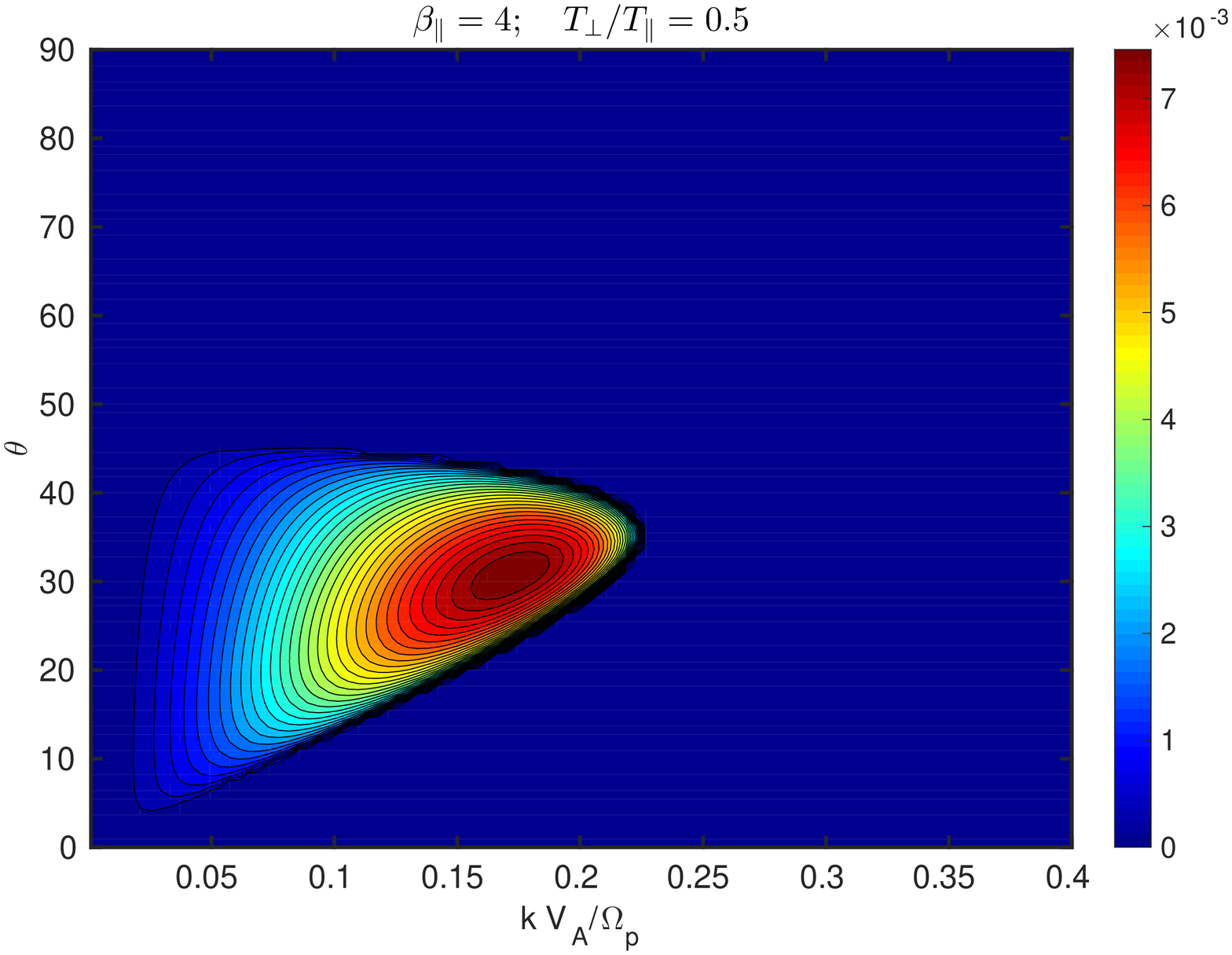}$$
  \caption{Solutions calculated exactly at the firehose threshold, $T_\perp/T_\parallel=0.5$, $\bpar=4$. Left: the Hall-CGL model.
    A strong instability exists at relatively large wavenumbers that is not observed in the kinetic solutions, as was also emphasized in Figure 9. 
    Right: the Hall-CGL-FLR1 model. The spurious instability at large wavenumbers is eliminated (not shown), and the FLR corrections weakly enhance 
    the oblique firehose instability at small wavenumbers. Since the solutions are calculated exactly at the firehose threshold, the right figure
    also implies that the Hall-CGL-FLR1 fluid solutions for the oblique firehose instability occur below the $k\to 0$ ``hard'' firehose threshold.} 
\end{figure*}

Nevertheless, we explored several plasma beta values from $\bpar=2.5$ to $\bpar=100$ and the results seem to be consistent across the entire range.
Close to the firehose threshold, the Hall-CGL-FLR1 model always yields a system with an oblique firehose instability. The contour plots for
$\bpar=2.5$ and $\bpar=100$ evaluated close to the firehose threshold are shown in Figure 17. Additionally, the case $\bpar=100$ is compared
in detail with kinetic solutions in Figure 18. The results are consistent with the case $\bpar=4$ that was discussed here in detail.
In Appendix B we investigate whether it is possible to neglect the Hall term for a sufficiently high plasma beta, i.e. $\bpar=100$,
since for such a high plasma beta the FLR corrections should be very strong. It turns out that in the case of oblique propagation close to the firehose threshold,
the answer is no. The Hall term can be neglected only for strictly parallel propagation. 
\begin{figure*}
$$\includegraphics[width=0.48\linewidth]{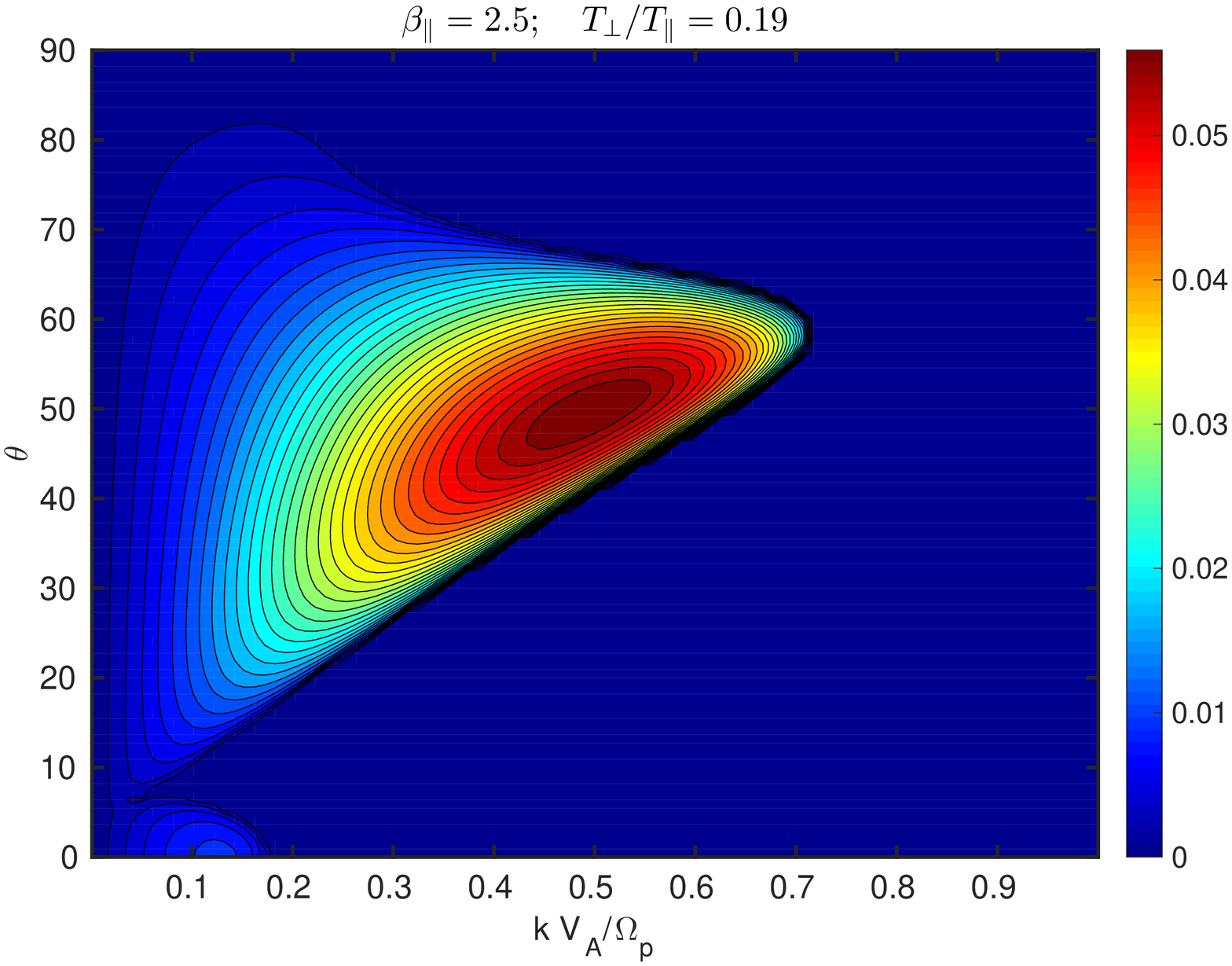}\hspace{0.03\textwidth}\includegraphics[width=0.48\linewidth]{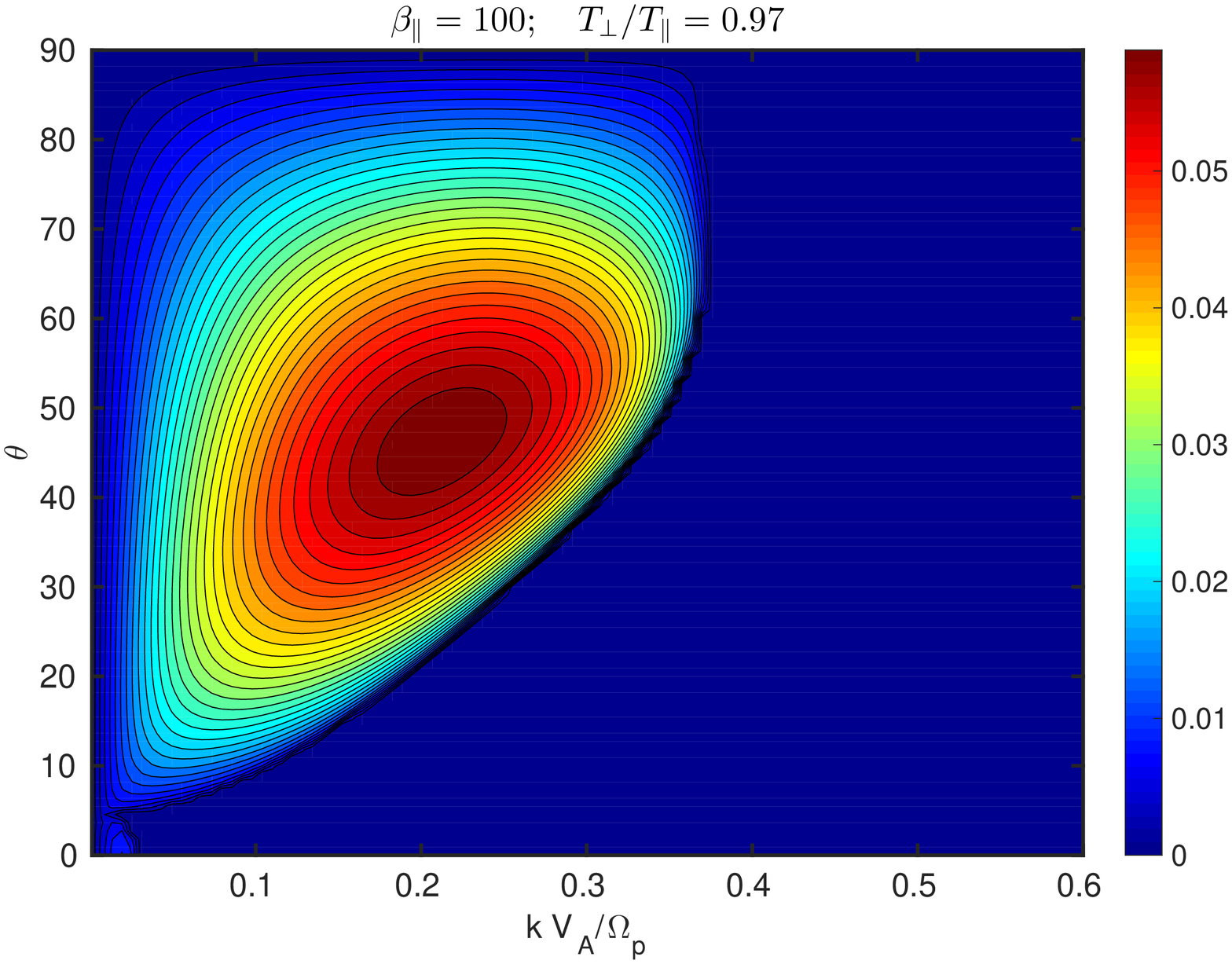}$$
  \caption{The maximum growth rate for different $\bpar$ values close to the corresponding firehose threshold. The Hall-CGL-FLR1 model is used.
    Left: $\bpar=2.5$, $T_\perp/T_\parallel=0.19$ (the threshold is $T_\perp/T_\parallel=0.2$). 
    Right: $\bpar=100$, $T_\perp/T_\parallel=0.97$ (the threshold is $T_\perp/T_\parallel=0.98$). In both cases, the parallel firehose instability is
    strongly suppressed.} 
\end{figure*}
\begin{figure*}
$$\includegraphics[width=0.48\linewidth]{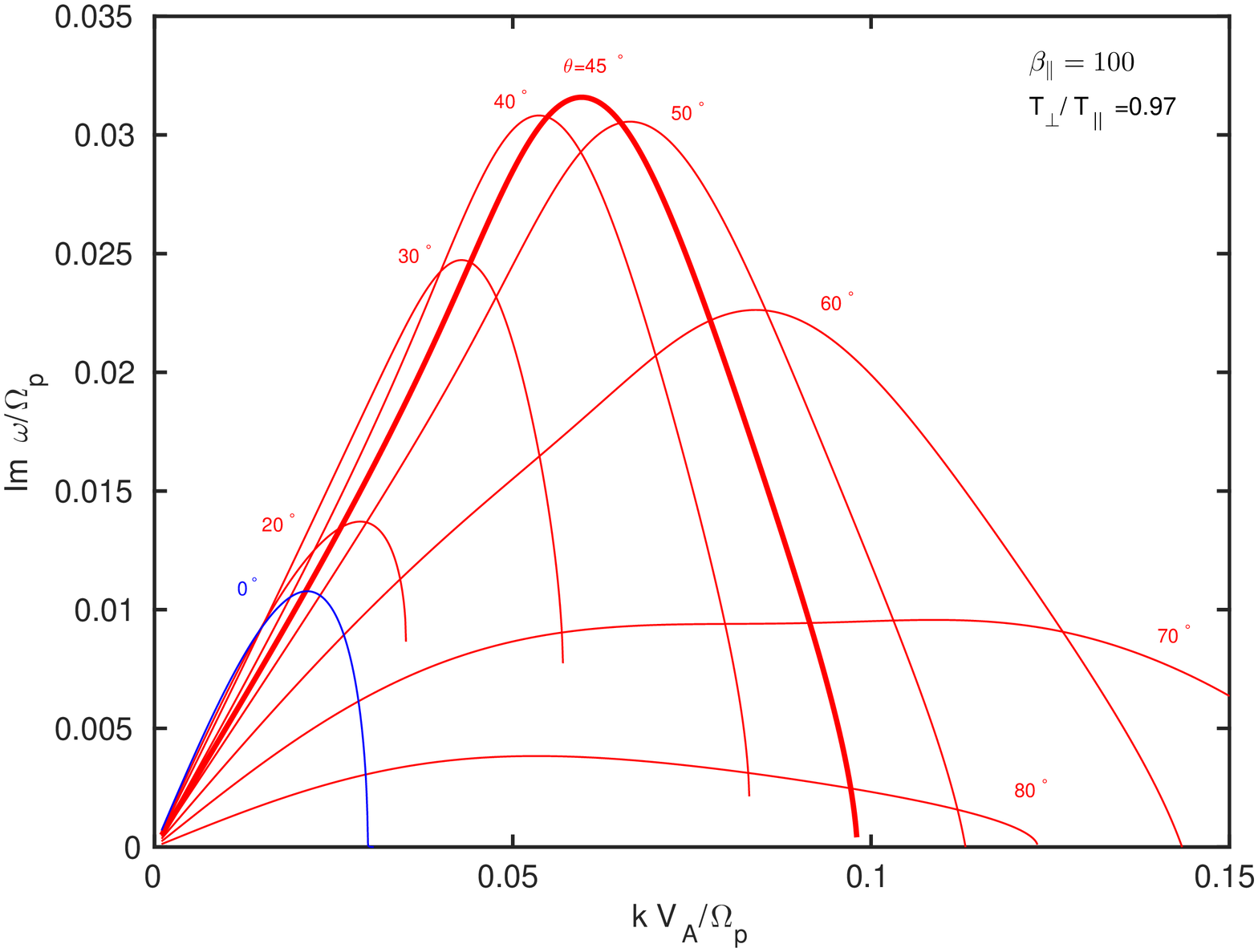}\hspace{0.03\textwidth}\includegraphics[width=0.48\linewidth]{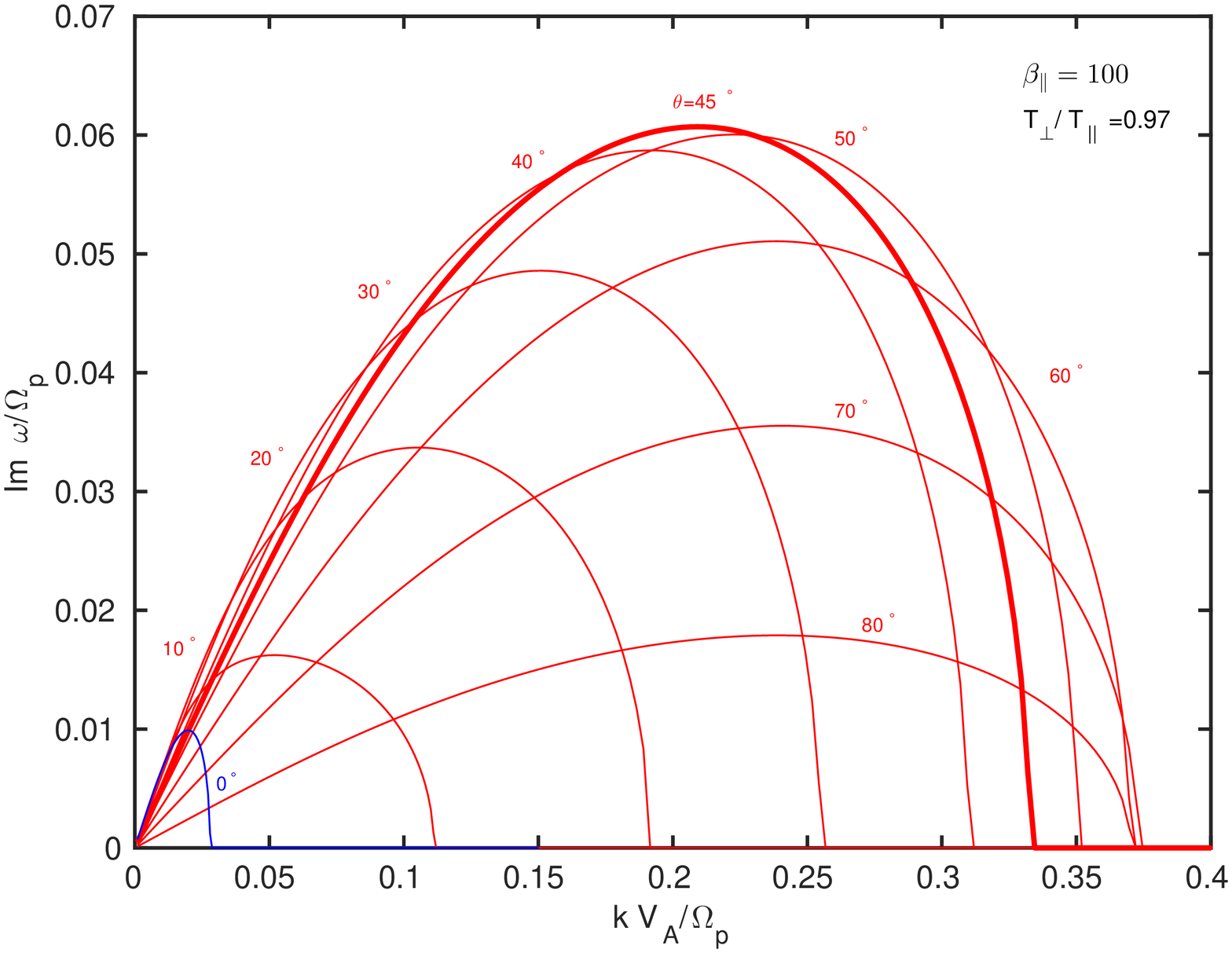}$$
  \caption{The angular dependence of the oblique firehose instability growth rate (red lines) for $\bpar=100$, $T_\perp/T_\parallel=0.97$.
    Left: kinetic solutions; Right: Hall-CGL-FLR1 solutions.
    In contrast to the parallel firehose instability with propagation angle $\theta=0^\circ$ (blue lines) where the growth rate is reproduced very accurately
    (see also Figure 8 right), the solutions for the oblique firehose instability are not reproduced with the same degree of precision, i.e., the kinetic
    solutions reach the maximum growth rate at smaller wavenumbers and the value of the maximum growth rate is smaller.  
    Nevertheless, both kinetic and fluid solutions reach a maximum growth rate around $\theta=45^\circ$, the maximum growth rate has an error of less than a
    factor of two ($\gamma_{max}=0.0316$ and $0.0607$), and the parallel firehose instability is strongly suppressed.} 
\end{figure*}

Finally, in Figure 19 we calculate marginally stable solutions with a prescribed maximum growth rate $\gamma_{max}$
for the oblique firehose instability in the $(\bpar,a_p)$ plane. The fitting parameters for the kinetic solutions were provided to us by Petr Hellinger
(Hellinger, private communication) and are represented by the red solid lines.
The dashed red lines are solutions of the Hall-CGL-FLR1 fluid model. The kinetic solution with $\gamma_{max}=10^{-3}$ is
almost indistinguishable from the solution with $\gamma_{max}=10^{-2}$, the same is true for the solutions of the fluid model, and only contours with $\gamma_{max}=10^{-2}$
and $\gamma_{max}=10^{-1}$ are plotted. The solutions are not trivial to calculate because the peak of the growth rate keeps moving in the $(k,\theta)$ plane and it is
not possible to prescribe a fixed wavenumber and angle of propagation. Instead, we prescribed $\omega/\Omega_p=i\gamma_{max}$ (since the oblique firehose is a non-propagating
instability) and for a given $\bpar$, we numerically found the maximum value of $a_p$ that lies on the implicitly given 3D surface in the variables $(k,\theta,a_p)$.
The numerical maximization procedure (we used MAPLE software) can be very sensitive to the initialization point and to the given range of variables, otherwise
no solution is found. To automatically calculate
solutions for different $\bpar$ values (over 300 points are plotted for each $\gamma_{max}$ fluid curve), the curve was separated into two (for $\gamma_{max}=10^{-1}$)
or four segments (for $\gamma_{max}=10^{-2}$) and special attention was paid to ensure that the maximum was reached inside the prescribed boundaries and not at the boundaries.
We note that for the smallest $\gamma_{max}=10^{-3}$, the procedure failed to converge for some $\bpar$ values. Nevertheless, the $\gamma_{max}=10^{-3}$ solution
is extremely close to the $\gamma_{max}=10^{-2}$ solution and only the latter is plotted. Figure 19 shows that fluid solutions do not match the kinetic contours,
but, the location in the $(\bpar,a_p)$ plane is approximately correct. This is an excellent result for such a simple fluid model. 

\begin{figure*}
$$\includegraphics[width=0.48\linewidth]{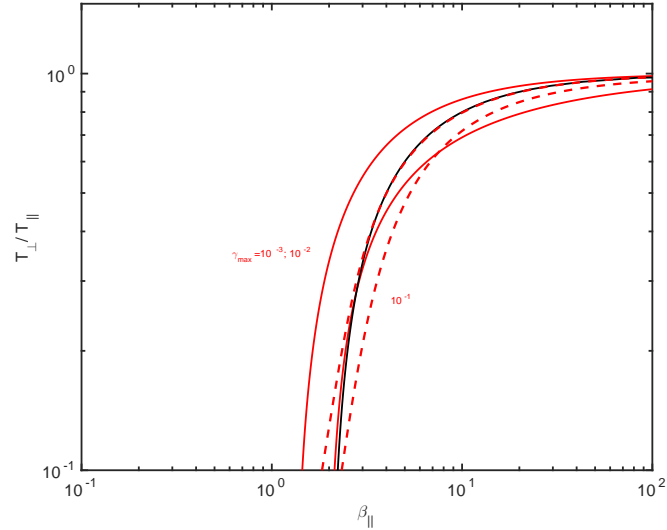}$$
  \caption{Marginally stable states for the oblique firehose instability with a prescribed $\gamma_{max}$.
    The solid lines are kinetic solutions and the dashed lines solutions of the Hall-CGL-FLR1 fluid model. Both in kinetic theory and in the fluid model,
    solutions with $\gamma_{max}=10^{-3}$ and $\gamma_{max}=10^{-2}$ are basically indistinguishable and only the latter is plotted. The kinetic contours are
    from \cite{Hellinger2006}, Figure 1 right (Hellinger, private communication). The (long-wavelength limit) hard firehose threshold is the black solid line. Note that the
    $10^{-2}$ (and $10^{-3}$) fluid contour is actually below the hard firehose threshold, which is nicely visible for the lowest $\bpar$ values and, the fluid contour
    reaches values slightly below $\bpar=2$.} 
\end{figure*}
\newpage
\section{The Myth of the CGL Description}
In concluding, we need to address an erroneous claim made frequently in the solar wind literature. This is that
CGL predicts extremely large temperature anisotropies in the radially expanding solar wind.
This misleading conclusion is arrived at by considering the steady-state CGL equations, i.e. neglecting the time derivatives, 
and considering only the evolution of mean quantities. Using the parallel and perpendicular temperatures instead of pressures,
the CGL equations in ``conservation'' form are written in steady-state as  
\begin{equation}
T_\parallel\sim \frac{\rho^2}{|\bb|^2}; \qquad T_\perp \sim |\bb|; \qquad \frac{T_\perp}{T_\parallel}=\frac{|\bb|^3}{\rho^2}. 
\end{equation}  
For a radially expanding solar wind, the density $\rho\sim 1/r^2$ with heliocentric distance $r$, 
and the radial or azimuthal magnetic field is either $|\bb|\sim 1/r^2$, or $|\bb|\sim 1/r$.
On choosing the radial magnetic field profile, we have 
\begin{equation} \label{eq:TempOne}
T_\parallel \sim \textrm{const.}; \qquad T_\perp \sim 1/r^2; \qquad \frac{T_\perp}{T_\parallel}\sim 1/r^2.
\end{equation}
The azimuthal magnetic field profile yields
\begin{equation} \label{eq:TempTwo}
T_\parallel \sim 1/r^2; \qquad T_\perp \sim 1/r; \qquad \frac{T_\perp}{T_\parallel}\sim r.
\end{equation}
There is also a curious case if the magnetic field profile is assumed to be $|\bb|\sim 1/r^{4/3}$, i.e. intermediate to
$1/r$ and $1/r^2$, which yields
\begin{equation} \label{eq:TempCrit}
T_\parallel \sim 1/r^{4/3}; \qquad T_\perp \sim 1/r^{4/3}; \qquad \frac{T_\perp}{T_\parallel}\sim \textrm{const.}.
\end{equation}
In (\ref{eq:TempCrit}), the temperature anisotropy stays constant. 
The results (\ref{eq:TempOne}) and (\ref{eq:TempTwo}) are often interpreted as that the CGL model yields extremely large temperature
anisotropies in the radially expanding solar wind. This simple interpretation is erroneous.
For $\rho\sim 1/r^2$, an interplanetary magnetic field (IMF) dependence steeper than $r^{-4/3}$ will evolve the system towards the firehose
threshold and an IMF dependence flatter than $r^{-4/3}$ towards the mirror threshold. Despite not exhibiting all the complexities
of the kinetic model, the CGL model captures both firehose instabilities properly at sufficiently long spatial scales, and both
instabilities are stabilized with the inclusion of the Hall term and FLR corrections.
Although the mirror instability was not addressed here, and despite the simplest mirror threshold being incorrect with respect to kinetic theory
(more elaborate models can rectify the difference with the inclusion of the heat flux), a threshold is indeed still present.   
Therefore, the CGL model will not allow large
temperature anisotropies in the expanding solar wind and, regardless of the dominant heating processes, the developing temperature anisotropies will be bounded,
if not completely by the same thresholds, just as the kinetic description is bounded. 

The same erroneous conclusion would be obtained if the time derivative $\pr_t$ were neglected in the Vlasov equation,
i.e. by considering kinetic theory in a steady-state. Instabilities are found from 
the dispersion relation and prescribing $\pr_t=0$ simply implies that the frequency $\omega=0$. 
The zero frequency condition can be useful in determining some thresholds for nonpropagating instabilities,
but one has to specifically derive these solutions from the dispersion relations.
Most importantly, instabilities refer to \emph{fluctuating} quantities. The fluctuations will become unstable - with a positive
growth rate - when the \emph{mean} values of these fluctuations satisfy some relations/thresholds. For example the oblique Alfv\'en wave (\ref{eq:ALF})
in the CGL description becomes unstable when the firehose threshold (\ref{eq:firehose}) is satisfied. 
It is the fluctuations that become unstable, and an analysis of mean/global quantities can not reveal an instability. The very simple calculations
(\ref{eq:TempOne}) and (\ref{eq:TempTwo}) above
therefore have to be approached with caution, since these calculations cannot reveal, by a definition, any instability.
So many papers in the solar wind literature erroneously claim that CGL predicts enormously large temperature anisotropies,
that it merits the phrase ``The Myth of the CGL Description''.

{\bf Acknowledgments} We would like to thank Thierry Passot and the referee, Alexander Schekochihin, for very useful discussions and suggestions.
We also thank Petr Hellinger for providing the kinetic contours that were used in Figures 7 \& 19.

\appendix
\section{The Hall-CGL-FLR1 dispersion relation}
The polynomial representing the FLR1 contributions in the Hall-CGL model can be written in the following form
\begin{eqnarray}
  \mathcal{P}^{\textrm{FLR}} &=& A_4' \omega^4 - A_2'\omega^2 +A_0';\\
  A_4' &=& \bpar^2\Big[ k_\parallel^4 \Big(1-\frac{a_p}{2}\Big)^2 +k_\perp^4\frac{a_p^2}{16}+k_\parallel^2 k_\perp^2 a_p \Big(1-\frac{a_p}{2}\Big) \Big];\\
  A_2' &=& \bpar k_\parallel^2 \Bigg\{ k_\parallel^6 \bpar \Big(1-\frac{a_p}{2}\Big)^2
  +k_\parallel^4 \Big[ \frac{3}{2}\bpar^2 \Big(1-\frac{a_p}{2} \Big)^2 +k_\perp^2\bpar\Big( 1-\frac{a_p^2}{4}\Big) +v_{A\parallel}^2(a_p-2) \Big]  \nn\\
  && +k_\parallel^2 k_\perp^2\Big[ -1+\frac{a_p}{2}+\bpar\frac{a_p}{4}\Big( 1-a_p+\bpar(1-\frac{a_p}{2}) \Big)
    -v_{A\parallel}^2\Big( 1-\bpar\frac{a_p}{2}(1-\frac{a_p}{2}) \Big) +k_\perp^2 a_p\bpar (1-\frac{7}{16}a_p) \Big]   \nn \\
  && +k_\perp^4 \frac{a_p}{4}\Big[  -1+2\bpar-v_{A\parallel}^2-a_p\bpar\Big( 1+\bpar(a_p-\frac{15}{8})-\frac{1}{4}k_\perp^2\Big)  \Big] \Bigg\};\\
  A_0' &=& \bpar^2 k_\parallel^4\Bigg\{ k_\parallel^6\frac{3}{2}\bpar\Big(1-\frac{a_p}{2}\Big)^2
  +k_\parallel^4 \Big(1-\frac{a_p}{2}\Big)\Big[ -3v_{A\parallel}^2+k_\perp^2\bpar\frac{3}{2}\Big(1-\frac{a_p}{2}+\frac{a_p^2}{6}\Big) \Big] \nn\\
  && +k_\parallel^2 k_\perp^2 \Big[ k_\perp^2\bpar a_p^2 (\frac{21}{32}-\frac{5}{16}a_p) -v_{A\parallel}^2\frac{3}{2}(1-\frac{a_p}{2}+\frac{1}{6}a_p^2 )
    -\Big(1-\frac{a_p}{2}\Big)\frac{3}{2}\Big( 1+\frac{1}{6}a_p\bpar-\frac{1}{6}a_p^2\bpar\Big) \Big] \nn\\
  && +k_\perp^4 a_p \Big[ k_\perp^2 a_p\bpar(\frac{13}{32}-\frac{3}{16}a_p) +\frac{v_{A\parallel}^2}{8}(1-2a_p) -\frac{1}{8}+\frac{3}{16}a_p\bpar(1-a_p)\Big] \Bigg\},
\end{eqnarray}  
where for simplicity we dropped the tilde symbol on all the quantities, including $\omega,k$ and $v_{A\parallel}^2=1+\frac{\bpar}{2}(a_p-1)$.
It is important to note that if the Hall term is neglected and only the FLR corrections are considered, i.e. if the CGL-FLR1 fluid is considered,
a different (and quite simpler) polynomial $\mathcal{P}^{\textrm{FLR}}$ is obtained, since the above polynomial contains coupled Hall and FLR contributions.
\section{The CGL-FLR1 model (no Hall term)}
The dispersion relation of the CGL-FLR1 fluid model reads (dropping the tilde symbol everywhere)
\begin{equation}
\Big(\omega^2-k_\parallel^2 v_{A\parallel}^2\Big)\Big(\omega^4-A_2\omega^2+A_0\Big) = \mathcal{P}^{\textrm{FLR-only}},
\end{equation}
where the CGL parameters $A_2,A_0$ are (\ref{eq:CGLparameters}) and the FLR1 contribution on the right hand side is
\begin{eqnarray}
  \mathcal{P}^{\textrm{FLR-only}} &=& \omega^2 \bpar^2 (A_2'' \omega^2 -A_0'');\\
  A_2'' &=& k_\parallel^4 \Big(1-\frac{a_p}{2}\Big)^2 +k_\perp^4\frac{a_p^2}{16}+k_\parallel^2 k_\perp^2 a_p \Big(1-\frac{a_p}{2}\Big);\\
  A_0'' &=& k_\parallel^2 \Bigg\{ k_\parallel^4 \frac{3}{2}\bpar \Big(1-\frac{a_p}{2}\Big)^2
  +k_\parallel^2 k_\perp^2 \frac{a_p}{2}\Big(v_{A\parallel}^2+\frac{\bpar}{2}\Big)\Big(1-\frac{a_p}{2}\Big)
  +k_\perp^4 \frac{a_p}{2}\Big[ 1-\frac{a_p}{2}+a_p\bpar \Big(\frac{15}{16}-\frac{a_p}{2}\Big) \Big] \Bigg\},
\end{eqnarray}
where again $v_{A\parallel}^2=1+\frac{\bpar}{2}(a_p-1)$. 
The dispersion relation for strictly parallel propagation ($k_\perp=0$) reads
\begin{equation}
\omega = \pm k_\parallel^2 \frac{\bpar}{2}\Big(1-\frac{a_p}{2}\Big) +k_\parallel \sqrt{v_{A\parallel}^2+\frac{k_\parallel^2\bpar^2}{4}\Big(1-\frac{a_p}{2}\Big)^2},
\end{equation}
where again a further two solutions are obtained by substituting $\omega$ with $-\omega$. For strictly parallel propagation, the
FLR corrections therefore stabilize the firehose instability and for example if Figure 8 is re-plotted with this model (not shown),
the solutions are identical to the green dashed curves obtained with the Hall-CGL-FLR1 model. 
However, this temptingly simple model fails for oblique propagation, since it does not stabilize the oblique firehose instability. 
In Figure B1 we plot the firehose growth rate when close to the firehose threshold for $\bpar=100$. The solutions are not stabilized and instead,
the growth rates converge to an asymptotic values. Figure B1 shows that the Hall term can not be neglected even for very large values of $\bpar$.
\begin{figure*}
$$\includegraphics[width=0.48\linewidth]{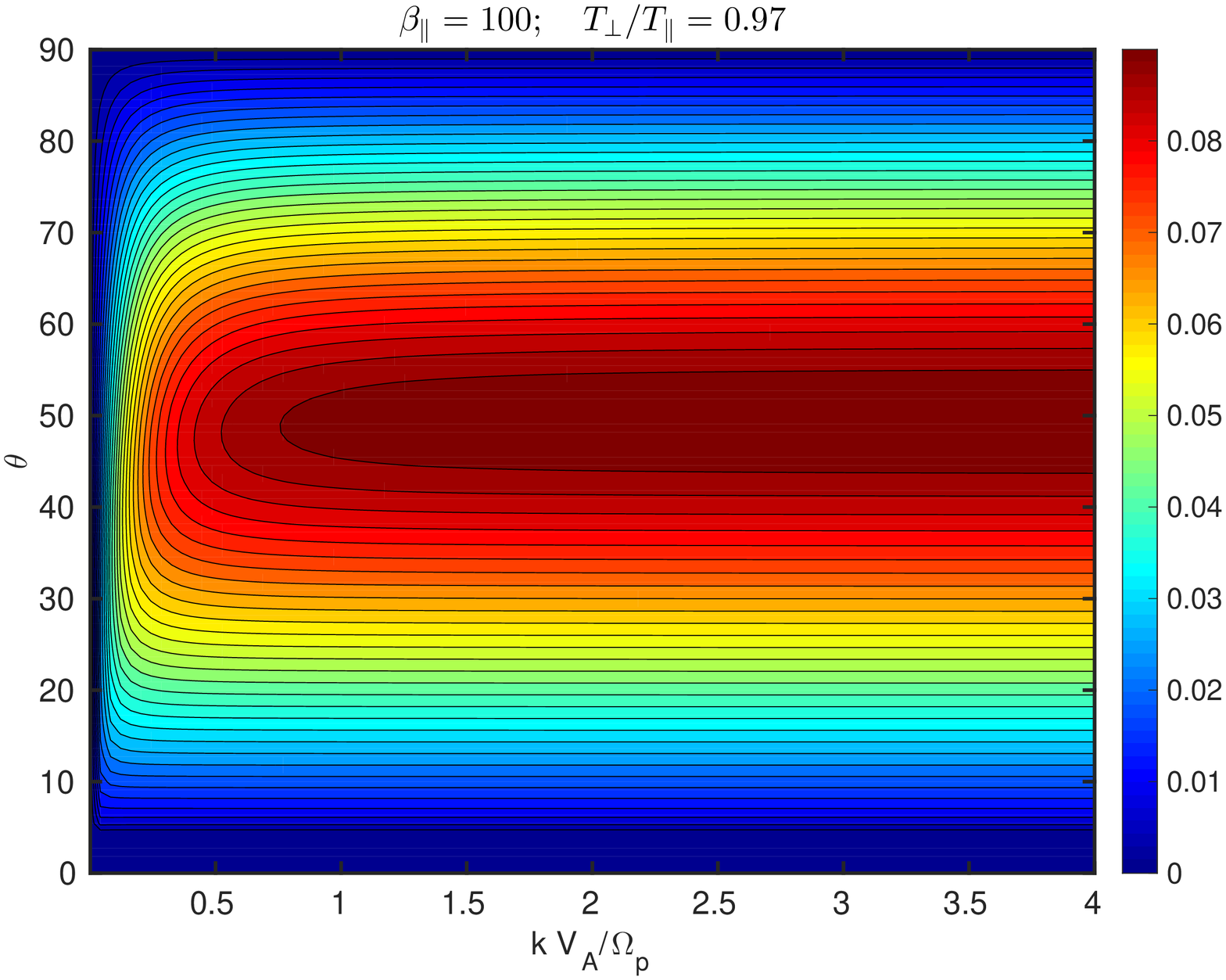}\hspace{0.03\textwidth}\includegraphics[width=0.48\linewidth]{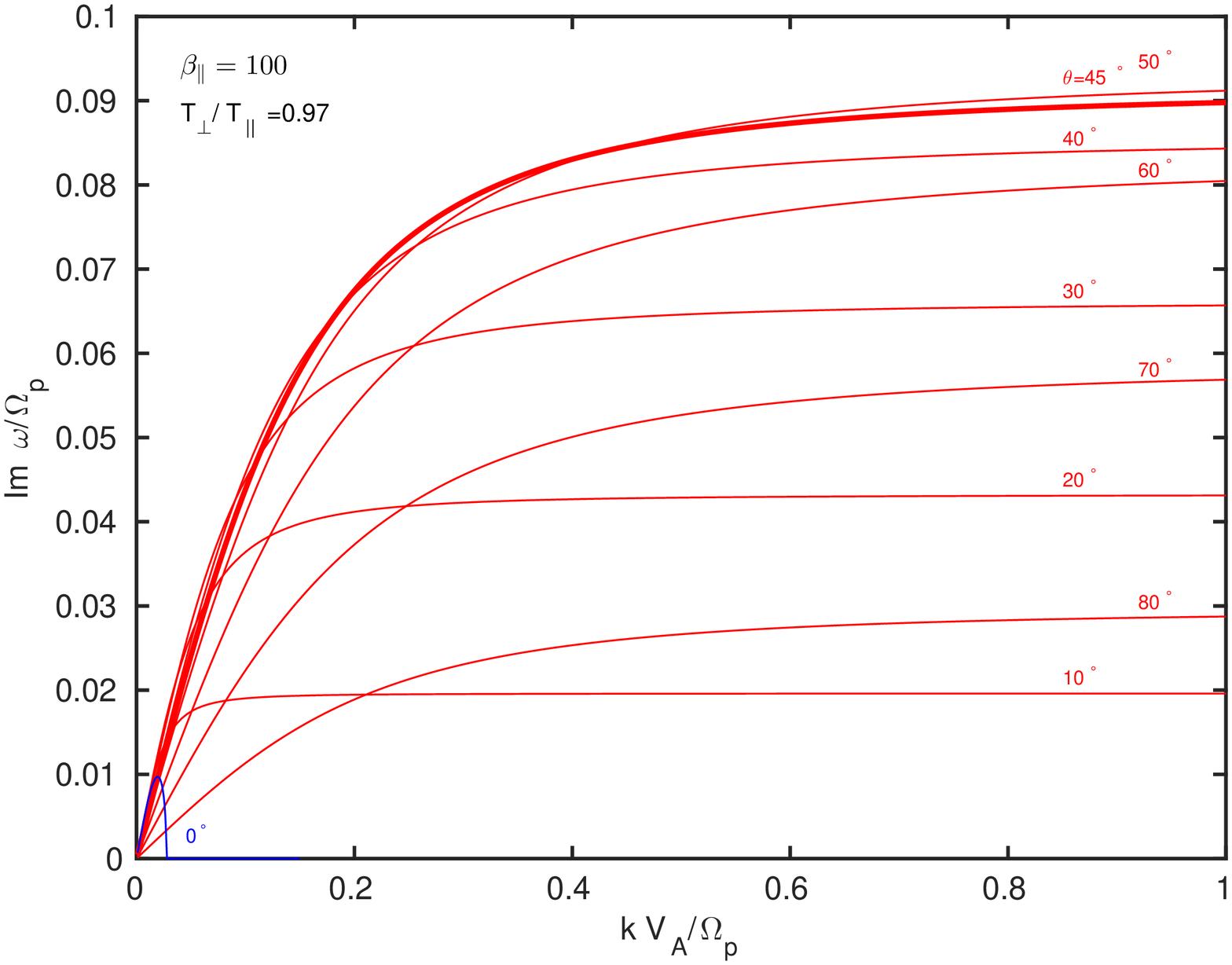}$$
  \caption{Solutions using the CGL-FLR1 model, i.e., the Hall term is turned off. Parameters are $\bpar=100$, $T_\perp/T_\parallel=0.97$ and
    the results should be compared with Figures 17 right \& 18. The solution with $\theta=45^\circ$ is emphasized with a thicker line, even though the
    solution with $\theta=50^\circ$ reaches a slightly higher maximum growth rate. The solutions are not stabilized at any range of
    wavenumbers, e.g., note that the left figure is plotted until $kV_A/\Omega_p=4$ and the solutions are not stabilized even beyond that.
    This implies that, for oblique propagation, the Hall term can not be neglected even for very high $\bpar$ values.} 
\end{figure*}



\hyphenation{Post-Script Sprin-ger}

\end{document}